\documentclass[aps,prb,twocolumn,superscriptaddress,floatfix,10pt]{revtex4-1}
\usepackage[utf8]{inputenc}

\usepackage{natbib}
\usepackage{graphicx}
\usepackage{latexsym}
\usepackage{amssymb}
\usepackage{amsmath}
\usepackage{amsfonts}
\usepackage{bm}
\usepackage{enumitem}
\usepackage{hyperref}
\usepackage{lipsum}
\usepackage{braket}
\usepackage{tikz}
\usepackage{comment}
\usetikzlibrary{arrows}
\usepackage[normalem]{ulem} 

\DeclareMathAlphabet\mathbfcal{OMS}{cmsy}{b}{n}
\newtheorem{nthm}{Theorem}[section]
\newtheorem{nlemma}[nthm]{Lemma}

\newcommand{\sfigref}[2]{Fig.\,\hyperref[#1]{\ref{#1}(#2)}}

\newcommand{\xc}[1]{{\color{red} #1}}
\newcommand{\xcc}[1]{{\color{orange} [(Xie) #1]}}
\newcommand{\xm}[1]{{\color{purple} #1}}
\newcommand{\xmc}[1]{{\color{blue} [#1]}}

\newcommand{\ws}[1]{{\color{cyan} #1}}

\newcommand{\ml}[1]{{\color{brown} (Michael) #1}}

\hypersetup{
  colorlinks = true,
  urlcolor = blue,
  pdfauthor = {Xiuqi Ma, Wilbur Shirley, Meng Cheng, Michael Levin, John McGreevy, Xie Chen},
  pdftitle = {Ma et al. - 2020 - Fractonic order in infinite-component Chern-Simons gauge theories}
}

\begin{document}

\title{Fractonic order in infinite-component Chern-Simons gauge theories}
\date{\today}

\author{Xiuqi Ma}
\affiliation{Department of Physics and Institute for Quantum Information and Matter, \mbox{California Institute of Technology, Pasadena, California 91125, USA}}

\author{Wilbur Shirley}
\affiliation{Department of Physics and Institute for Quantum Information and Matter, \mbox{California Institute of Technology, Pasadena, California 91125, USA}}

\author{Meng Cheng}
\affiliation{\mbox{Department of Physics, Yale University, New Haven, CT 06520-8120, USA}}

\author{Michael Levin}
\affiliation{Department of Physics, James Franck Institute, University of Chicago, Chicago, IL 60637, USA}

\author{John McGreevy}
\affiliation{Department of Physics, University of California at San Diego, La Jolla, California 92093, USA}

\author{Xie Chen}
\affiliation{Department of Physics and Institute for Quantum Information and Matter, \mbox{California Institute of Technology, Pasadena, California 91125, USA}}

\begin{abstract} $2+1$D multi-component $U(1)$ gauge theories with a Chern-Simons (CS) term provide a simple and complete characterization of $2+1$D Abelian topological orders. In this paper, we extend the theory by taking the number of component gauge fields to infinity and find that they can describe interesting types of $3+1$D ``fractonic'' order. ``Fractonic'' describes the peculiar phenomena that point excitations in certain strongly interacting systems either cannot move at all or are only allowed to move in a lower dimensional sub-manifold. In the simplest cases of infinite-component CS gauge theory, different components do not couple to each other and the theory describes a decoupled stack of $2+1$D fractional Quantum Hall systems with quasi-particles moving only in 2D planes --- hence a fractonic system. We find that when the component gauge fields do couple through the CS term, more varieties of fractonic orders are possible. For example, they may describe foliated fractonic systems for which increasing the system size requires insertion of nontrivial $2+1$D topological states. Moreover, we find examples which lie beyond the foliation framework, characterized by 2D excitations of infinite order and braiding statistics that are not strictly local.
\end{abstract}

\maketitle

\section{Introduction}
\label{sec:intro}


Fracton models \cite{Nandkishore_2019, Pretko_2020} are characterized by the peculiar feature that some of their gapped point excitations are completely localized or are restricted to move only in a lower dimensional sub-manifold. Two large classes of models have been studied extensively, with very different features. The exactly solvable  fracton models (see for example \onlinecite{ChamonModel,HaahCode, Sagar16, VijayFracton,Prem_2019, Song2019}),
on the one hand, are gapped and exhibit properties like exponential ground state degeneracy, nontrivial entanglement features, \cite{HermeleEntropy, ShiEntropy, BernevigEntropy, Shirley_2019} and foliation structure. \cite{Shirley_2018} The higher rank continuum gauge theories (see for example \onlinecite{PretkoU1, electromagnetismPretko, Gromov2019, seiberg2020exotic}), on the other hand, host gapless photon excitations, on top of which gapped fracton excitations emerge due to nontrivial forms of symmetry like dipole conservation. The fracton models discovered so far host features that are very similar to those in topological models like fractional quantum Hall and (rank-1) gauge theories, but also generalize the topological framework in nontrivial ways.


One theoretical tool that plays an important role in the study of $2+1$D topological phases is Chern-Simons gauge theory. \cite{WenRigid} In particular, it has been shown that multi-component $U(1)$ gauge theories with a Chern-Simons term give a complete characterization of $2+1$D abelian topological phases. \cite{WenChernSimons} The Lagrangian of the theory is given by
\begin{equation}
\label{eq:lagrangian}
\mathcal{L} = -\frac{1}{4e^2}\sum_i \mathcal{F}^i_{\mu\nu}\mathcal{F}^{i,\mu\nu} + \frac{1}{4\pi} \sum_{ij} K_{ij} \epsilon^{\mu\nu\lambda} \mathcal{A}^i_{\mu}\partial_{\nu}\mathcal{A}^j_{\lambda},
\end{equation}
where $\mu,\nu,\lambda = 0,1,2$, and $i,j$ label the different gauge fields and take values in a finite set $i,j=1,...,N$. The matrix $K$ is an $N\times N$ symmetric integer matrix. The universal topological features are captured in the $K$ matrix, from which one can derive the ground state degeneracy, anyon fusion and braiding statistics, edge states, etc. of the topological phase.\cite{WenChernSimons}

Can we take the number of gauge fields to infinity and extend this formalism to describe $3+1$D fractonic order? In this paper, we call such theories ``iCS'' theories, ``i'' for infinite. This idea comes from the simple observation that if we do take this extension and choose the infinite dimensional $K$ matrix to be simply diagonal (with diagonal entries being, for example, $3$),
\begin{equation}
K=
\begin{pmatrix}
\ddots\\
& 3\\
&  &  3\\
&  &  & 3\\
&  &  &  & \ddots
\end{pmatrix},\label{eq:Kd3}
\end{equation}
then the Lagrangian describes a decoupled stack of $2+1$D fractional quantum Hall states (each with filling fraction $\nu=1/3$ in this example). Such decoupled stacks of $2+1$D topological states, while simple, contain several of the key features of fracton physics: ground state degeneracy that increases exponentially with the height of the stack; anyon excitations that are mobile in 2D planes only and cannot hop vertically; and entanglement entropy of sub-regions that contains a sub-leading term which scales linearly with the height of the region.\cite{HermeleEntropy} Therefore, this simple stack system described by a diagonal infinite $K$ matrix is a fracton model, although a very trivial one. In the following discussion, we will always use the convention that each gauge field has $x$ and $y$ spatial components, but not a $z$ one, and we will try to interpret the $i$ index as a $z$ direction spatial coordinate (we will discuss the condition under which we can safely do so). 

Can iCS theories with more complicated $K$ matrices, e.g. quasi-diagonal ones, lead to more interesting types of fractonic behavior? In this paper, we show that this is indeed the case. In section~\ref{sec:gapped_foliated}, we show that some gapped iCS theories have foliated fractonic order, which was first identified in several exactly solvable fracton models. \cite{Shirley_2018, Shirley_2019, Shirley2019twisted} The iCS models cover both twisted and non-twisted foliated fractonic phases and can represent foliated phases without an exactly solvable limit. More interestingly, in section~\ref{sec:gapped_non-foliated}, we present a gapped iCS theory that is qualitatively different from any exactly solvable fracton model we know before. The ground state degeneracy does not follow exactly a simple exponential form, but approaches one in the thermodynamic limit. Quasiparticles move in planes and braiding statistics become ever more fractionalized as system size increases. This example represents a new class of gapped fractonic order. Note that the gapped iCS theories discussed in this paper are ``fractonic'' in the sense that they contain point ``planon'' excitations that move in 2D planes only but not the third direction. There are no true ``fracton'' excitations in these models which are completely localized when on their own. Finally, in section~\ref{sec:gapless}, we discuss an iCS theory which is gapless. On top of the gapless photon excitation, the system has a constant ground state degeneracy and fractional excitations generated by membrane operators. What kind of $3+1$D physics this model describes is an intriguing question. Some of these iCS theories have been studied in the context of three dimensional quantum Hall systems\cite{qiu1989phases,Qiu90-2, Naud00, Naud01} where their unusual properties of braiding statistics, edge states, etc., were first pointed out.

To substantiate the results we obtain from field theory analysis, we present an explicit lattice construction in section~\ref{sec:lattice}. The construction works for any quasi-diagonal infinite dimensional $K$ matrix with bounded entries (defined in section~\ref{sec:lattice}) and demonstrates that the corresponding iCS theory indeed describes the effective low energy physics of an anomaly-free $3+1$D local model. In particular, we write down a lattice Hamiltonian and the lattice form of the string operators for the planons, and calculate the spectrum (from field theory). 

Finally, in section~\ref{sec:conclusion}, we summarize our result and discuss the various open questions that follow the initial exploration of iCS theory presented here.

In Appendix~\ref{sec:Km_statistics} we discuss the tangential problem of how to construct the $K$ matrix representation of a $2+1$D abelian topological order if we are given the fusion group and statistics of its anyons. This translates to the math problem of quadratic forms on finite abelian groups and a complete solution is known. \cite{WALL1963281, wang2020abelian} We present the procedure step by step for interested physics readers.

Note that in this paper, we are considering $2+1$D gauge fields with $2+1$D gauge symmetries although the model is a $3+1$D model. It is possible to add a $z$ component to the gauge field and modify the model so that it satisfies $3+1$D gauge symmetries. We find that in most such cases, the model becomes gapless, similar to the case studied in \onlinecite{Levin2009}. We leave these cases out of the scope of this paper.

\section{Gapped foliated theories}
\label{sec:gapped_foliated}

\begin{figure}[tbp]
    \hbox{\hspace{-1em}
    \includegraphics[width=0.5\textwidth]{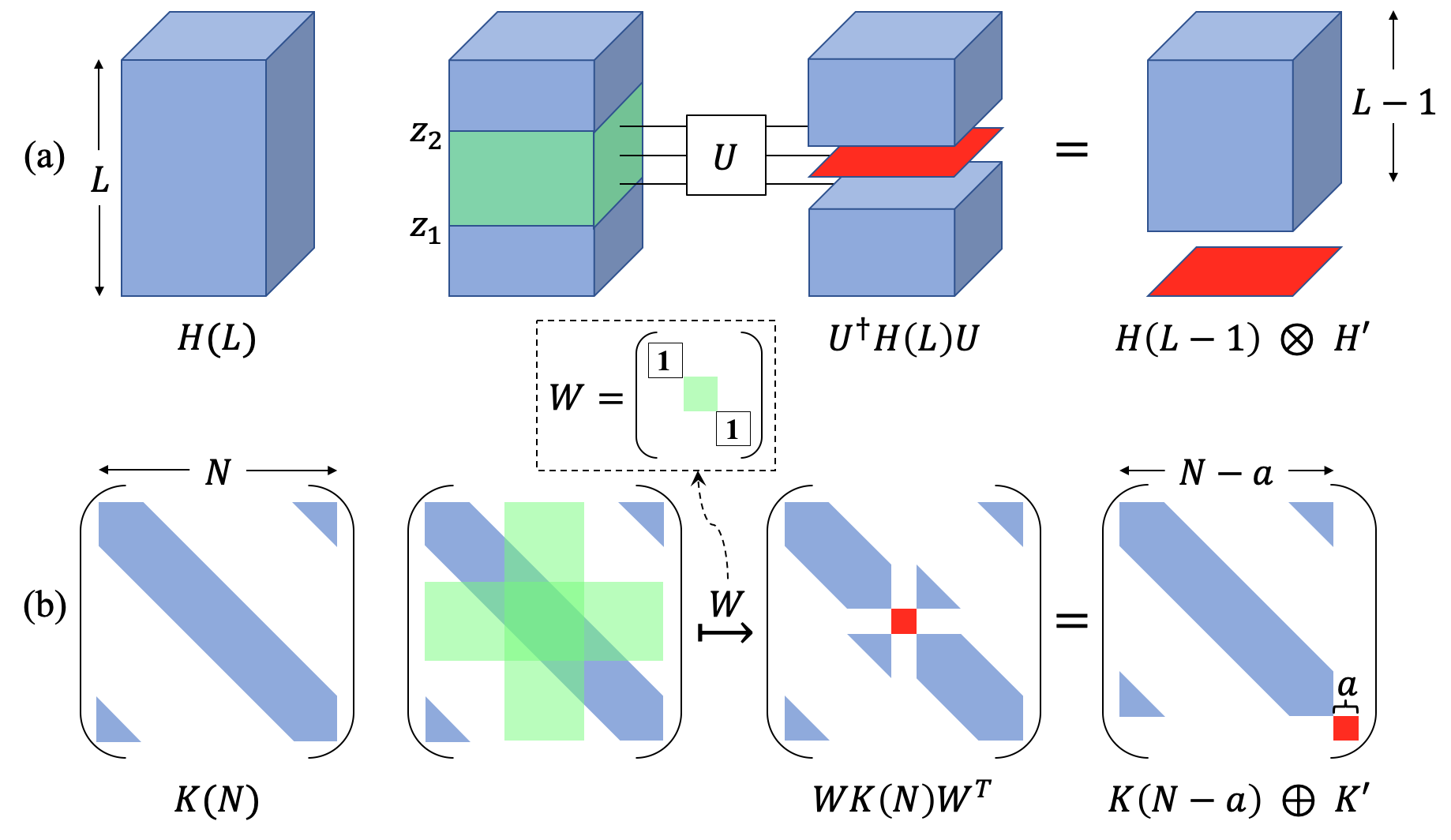}}
    \caption{Foliated fracton order and its interpretation in terms of $K$ matrix. In Figure~(a) (first line), we start with a system $H(L)$ of size $L$ in the $z$ direction. A finite depth local unitary circuit $U$ is applied to the green region $\{(x,y,z):z_1\le z\le z_2\}$. The result is the same system $H(L-1)$ of size $L-1$ in the $z$ direction and a decoupled $2+1$D gapped system (red layer). In Figure~(b) (second line), we start with a quasi-diagonal $K(N)$ of size $N\propto L$ with periodic boundary condition. Only entries in the blue region can be non-zero. We apply the transformation $K(N)\mapsto WK(N)W^T$, where $W\in\text{GL}(N,\mathbb{Z})$ shown in the dashed box is equal to the identity except in the green block, so the action of $W$ on $K(N)$ is within the green cross in the second figure. The result is the direct sum of the same system $K(N-a)$ of size $N-a$ and a decoupled block $K'$ of size $a=O(1)$ (red block).}
    \label{fig:foliation}
\end{figure}

A number of the fracton models discovered so far have a ``foliation structure''.\cite{Shirley_2018, Shirley_2019, Shirley2019twisted} That is, a model with a larger system size can be mapped under a finite depth local unitary circuit $U$ to the same model with a smaller system size together with decoupled layers of $2+1$D gapped states, as shown in Fig.~\ref{fig:foliation}~(a). For example, it was shown \cite{Shirley_2018} that the X-cube model of size $L_x\times L_y \times L_z$ can be mapped to one with size $L_x \times L_y \times (L_z-1)$ together with a $2+1$D toric code. Actually, the same process can be implemented in all $x$, $y$ and $z$ directions, and hence the X-cube model is said to be ``3-foliated''. Other fracton models with a ``foliation structure'' include the semionic X-cube model,\cite{MaLayers, Shirley_2019_excitation} the checkerboard model,\cite{Shirley_2019_checkerboard} the Majorana checkerboard model\cite{Wang_2019}.

An iCS theory can have a ``foliation structure'' as well and the $K$ matrix formulation provides a particularly simple mathematical framework to study it, as explained in Fig.~\ref{fig:foliation}~(b). Obviously the diagonal $K$ matrix, for example the one in Eq.~\ref{eq:Kd3}, represents a rather trivial 1-foliated fracton model where a model of height $L$ (in the stack direction) is the same as a model of height $L-1$ together with a decoupled 2D layer. Moreover, in Ref.~\onlinecite{Shirley2019twisted}, it was shown that it is possible to represent more nontrivial types of foliated fracton order using an iCS theory. In particular, it was shown that an infinite dimensional $K$ matrix of the form 
\begin{equation}
\label{eq:K_F}
\let\quad\; 
\def\-{\raisebox{.75pt}{-}} 
K_\text{F}=\bordermatrix{
     &&e_1&m_1&e_2&m_2&e_3&m_3&e_4&  \cr
&\ddots &  &  &  &  &  &  &  &  \cr
     && 0& 2& \-1&  &  &  &  &  \cr
     && 2& 0&  &  &  &  &  &  \cr
     && \-1&  & 0& 2& \-1&  &  &  \cr
     &&  &  & 2& 0&  &  &  &  \cr
     &&  &  & \-1&  & 0& 2& \-1&  \cr
     &&  &  &  &  & 2& 0&  &  \cr
     &&  &  &  &  &\-1 &  & 0&  \cr
     &&  &  &  &  &  &  &  & \ddots
}.
\end{equation}
describes a twisted 1-foliated fracton order. All non-zero entries in the matrix lie within distance $2$ from the main diagonal; the matrix is hence said to be quasi-diagonal. It is translation invariant with a period of $2$: $i\mapsto i+2$, $j\mapsto j+2$. We have added a subscript ``F'' to indicate that it is foliated. The meaning of the column labels will become clear once we take the inverse of this matrix.

To see what kind of physics this $K_\text{F}$ matrix describes, we first notice that the determinant of the $K_\text{F}$ matrix of size $2L$ is given by $\det{K_\text{F}(2L)} = (-4)^L$. Therefore, the ground state degeneracy on a 3D torus is given by
\[\log_2 \text{GSD} = 2L,\]
which takes a simple linear form in $L$. Next, the quasi-particle content can be read from the $K^{-1}_\text{F}$ matrix
\begin{equation}
\let\quad\;
K_\text{F}^{-1}=\frac{1}{4}\bordermatrix{
     && m_0&e_1&m_1&e_2&m_2&e_3&m_3&  \cr
&\ddots &  &  &  &  &  &  &  &  \cr
 && 0 &  & 1 &  &  &  &  &  \cr
 &&  & 0 & 2 &  &  &  &  &  \cr
 && 1 & 2 & 0 &  & 1 &  &  &  \cr
 &&  &  &  & 0 & 2 &  &  &  \cr
 &&  &  & 1 & 2 & 0 &  & 1 &  \cr
 &&  &  &  &  &  & 0 & 2 &  \cr
 &&  &  &  &  & 1 & 2 & 0 &  \cr
 &&  &  &  &  &  &  &  & \ddots 
}. \label{eq:KFi}
\end{equation}
The column labels $e_i$ and $m_i$ follow from those in Eq.~\ref{eq:K_F}. It is now easy to see that we choose these labels because the statistics of $e_i$ and $m_i$ are similar to those in a $\mathbb{Z}_2$ gauge theory where the $e$ and $m$ excitations are bosons and have a mutual $-1$ braiding statistic. But this $K_\text{F}$ matrix represents not just a decoupled stack of $\mathbb{Z}_2$ gauge theories, because the $m$ excitations have mutual $i$ statistics between neighbors. Indeed, it was shown in Ref.~\onlinecite{Shirley2019twisted} to describe a twisted 1-foliated fractonic order. That is
\begin{enumerate}
    \item The model is gapped and has fractional excitations that move only in the $xy$ plane, hence a fracton model.
    \item The model of height $L$ in the $z$ direction (corresponding to a $K_\text{F}$ matrix of size $4L$) can be mapped to one of height $L-1$ (corresponding to a $K_\text{F}$ matrix of size $4L-4$) together with a $2+1$D topological state layer (a twisted $\mathbb{Z}_2\times \mathbb{Z}_2$ gauge theory in this case).
    \item The model is not equivalent to a pure stack of $2+1$D topological layers. Note that the entries in $K_\text{F}^{-1}$ are strictly zero once we move sufficiently far away from the main diagonal.
\end{enumerate}
Comparing this to examples discussed in later sections, we see that this is a hallmark of foliated iCS theories.

The way to see the foliation structure is to apply a local, general linear transformation $W$ of the form
\begin{equation}
\let\quad\;
\def\-{\raisebox{.75pt}{-}}
W=\bordermatrix{
 && \tilde{e}_1&\tilde{m}_1&\tilde{e}^A&\tilde{m}^A&\tilde{e}^B&\tilde{m}^B&\tilde{e}_2&\tilde{m}_2&\tilde{e}_3&   \cr
&\ddots &  &  &  &  &  &  &  &  & &  \cr
e_1&& 1 &  &  &  & \-1 &  &  & \-1 & &  \cr
m_1&&  & 1 &  &  &  &  &  &  & &  \cr
e_2&&  &  & 1 &  &  &  &  &  & &  \cr
m_2&&  &  &  & 1 &  &  &  & 1 & &  \cr
e_3&&  &  &  &  & 1 &  &  &  & &  \cr
m_3&&  & 1 &  &  &  & 1 &  &  & &  \cr
e_4&&  &  & \-1 &  &  &  & 1 &  & &  \cr
m_4&&  &  &  &  &  &  &  & 1 & &  \cr
e_5&&  &  &  &  &  &  &  &  & 1 & \cr
 &&  &  &  &  &  &  &  &  & & \ddots 
}, \label{eq:W}
\end{equation}
so that $K_\text{F}$ is transformed into
\begin{equation*}
\let\quad\;
\def\-{\raisebox{.75pt}{-}}
WK_\text{F}W^T=\bordermatrix{
 && \tilde{e}_1&\tilde{m}_1&\tilde{e}^A&\tilde{m}^A&\tilde{e}^B&\tilde{m}^B&\tilde{e}_2&\tilde{m}_2&\tilde{e}_3&  \cr
&\ddots &  &  &  &  &  &  &  &  &  &  \cr
 && 0 & 2 &  &  &  &  & \-1 &  &  &  \cr
 && 2 & 0 &  &  &  &  &  &  &  &  \cr
 &&  &  & 0 & 2 & \-1 & 0 &  &  &  &  \cr
 &&  &  & 2 & 0 & 0 & 0 &  &  &  &  \cr
 &&  &  & \-1 & 0 & 0 & 2 &  &  &  &  \cr
 &&  &  & 0 & 0 & 2 & 0 &  &  &  &  \cr
 && \-1 &  &  &  &  &  & 0 & 2 & \-1 &  \cr
 &&  &  &  &  &  &  & 2 & 0 &  &  \cr
 &&  &  &  &  &  &  & \-1 &  & 0 &  \cr
 &&  &  &  &  &  &  &  &  &  & \ddots
},
\end{equation*}
where the middle $4\times 4$ block is decoupled from the rest of the system and the remaining part of the transformed $K$ matrix is the same as the original one in Eq.~\ref{eq:K_F}, only slightly smaller. Note that, although the $W$ matrix looks quite big, it acts non-trivially only within the finite block shown in Eq.~\ref{eq:W}. Its action is the identity outside. This transformation hence realizes the renormalization group transformation \cite{VidalRG, Shirley_2018} of the 1-foliated fracton model formulated in terms of infinite dimensional $K$ matrices, as shown schematically in Fig.~\ref{fig:foliation}~(b). 

The iCS theory, and correspondingly the infinite dimensional $K$ matrix, hence provide a convenient formulation for studying the foliation structure in a 1-foliated fracton model. The example discussed above can be generalized to a whole class of 1-foliated models with a similar foliation structure, as discussed in Appendix~\ref{sec:KFnr}.

\section{Gapped non-foliated theories}
\label{sec:gapped_non-foliated}

While the iCS formulation is useful in the study of foliated fracton models, a more surprising finding is that iCS theories can also be non-foliated. Among all type~I fracton models -- ones with mobile fractional excitations -- that we know so far, the abelian ones are all foliated. The iCS theory, being an abelian type~I fracton model, hence extends our understanding of what is possible within the realm of fractonic order.

Consider the iCS theory with a simple tridiagonal $K$ matrix
\begin{equation}
    \label{eq:KnF}
    K_{\text{nF}}=\begin{pmatrix}
    3 & 1 & & & 1 \\
    1 & 3 & 1 & & \\
    & \ddots & \ddots & \ddots & \\
    & & 1 & 3 & 1 \\
    1 & & & 1 & 3
    \end{pmatrix}.
\end{equation}
Note that we have taken periodic boundary condition in the matrix. The ``nF'' subscript denotes non-foliated. This theory was studied in Refs.~\onlinecite{Qiu90,Qiu90-2,Naud00,Naud01} as an effective theory for coupled fractional quantum Hall layers. Many aspects of its properties have been studied. Here we look at the theory from a fracton perspective, that is, to address the question: is this a fracton model and if so, what type of fracton model?

A field theory calculation shows that this theory is gapped (see section~\ref{sec:spectrum}). The determinant $D(N)$ of the matrix of size $N$, and hence the ground state degeneracy of the model on a 3D torus of height $N$, follow a rather complicated form
\begin{equation}
\label{eq:det(K)}
    D(N) = \left(\frac{3+\sqrt{5}}{2}\right)^N+\left(\frac{3-\sqrt{5}}{2}\right)^N-2(-1)^N.
\end{equation}
The exponential growth of GSD in system size indicates fractonic order. However, unlike in the foliated case, the GSD does not follow a simple exponential form (with possible pre-factors), but only approaches such a form in the thermodynamic limit $N\to \infty$ with an irrational base $\left(3+\sqrt{5}\right)/2$. 

Another way to see that this model is ``weird'' is from the fusion group and statistics of its planons. Such information can be read from the inverse of the matrix, which for size $N=5$ takes the form
\[K_{\text{nF}}^{-1} = \frac{1}{25} \begin{pmatrix} 11 & -4 & 1 & 1 & -4 \\ -4 & 11 & -4 & 1 & 1 \\ 1 & -4 & 11 & -4 & 1 \\ 1 & 1 & -4 & 11 & -4 \\ -4 & 1 & 1 & -4 & 11 \end{pmatrix},\]
and for size $N=7$ takes the form\ws{\footnote{The integers 1, 4, 11, 29, etc. form a sequence known as the Lucas numbers.}}
\[K_{\text{nF}}^{-1} = \frac{1}{65} \begin{pmatrix} 29 & -11 & 4 & -1 & -1 & 4 & -11 \\ -11 & 29 & -11 & 4 & -1 & -1 & 4 \\ 4 & -11 & 29 & -11 & 4 & -1 & -1 \\ -1 & 4 & -11 & 29 & -11 & 4 & -1 \\ -1 & -1 & 4 & -11 & 29 & -11 & 4 \\ 4 & -1 & -1 & 4 & -11 & 29 & -11 \\ -11 & 4 & -1 & -1 & 54 & -11 & 29 \end{pmatrix}.\]
Note the difference from the foliated case (for example Eq.~\ref{eq:KFi}). First of all, the magnitude of the entries decay exponentially away from the main diagonal, but they never become exactly zero. Secondly, each entry varies as the system size increases and approaches an irrational number as the system size goes to infinity
\begin{equation}
\label{eq:K_nF_inv}
    \left(K^{-1}_{\text{nF}}\right)_{ij} \to \frac{(-1)^{i-j}}{\sqrt{5}}\left(\frac{3+\sqrt{5}}{2}\right)^{-|i-j|}.
\end{equation}
In terms of quantum Hall physics, this indicates an irrational amount of charge in layer $j$ attached to a flux inserted in layer $i$.\cite{qiu1989phases} In terms of abelian topological order, this indicates an irrational phase angle in the braiding statistics between the $i$th anyon and the $j$th anyon
\begin{equation*}
    \theta_{ij} = 2\pi\frac{(-1)^{i-j}}{\sqrt{5}}\left(\frac{3+\sqrt{5}}{2}\right)^{-|i-j|}.
\end{equation*}
$K_{\text{nF}}$ of size $N$ gives a fusion group $G_N = \mathbb{Z}_{F_N} \times \mathbb{Z}_{5F_N}$, where $F_N$ is the $N$th number in the Fibonacci sequence. Therefore, the fusion group has two generators, one of order $F_N$, the other of order $5F_N$. 

These features preclude a foliation structure in $K_{\text{nF}}$. In both cases, the fusion group is exponentially large and correspondingly the ground state degeneracy grows exponentially with system size. But the underlying reasons for this growth are very different. In the foliated models, as the planons come from the hidden 2D layers, they have finite orders and correspondingly rational statistics. At the same time, the fusion group has a lot of generators, a number that grows linearly with system size. In the non-foliated example however, the fusion group has only two generators, each of infinite order (exponentially growing with system size). Their self and mutual statistics also become more and more fractionalized as the system size grows and eventually approach an irrational number. 

It is therefore straightforward to see that the theory represented by $K_{\text{nF}}$ cannot be foliated. In particular, not every local (in the $z$ direction) planon can be decoupled into an anyon in a foliation layer. First, the ground state degeneracy does not follow a simple formula of $ab^N$, with $b$ being an integer or a root of an integer, as expected in a foliated model. Secondly, the elementary planon has an infinite order and nontrivial (although exponentially decaying) statistics with planons a large distance away. This cannot happen in a foliated model. In a foliated model, when each foliation layer is inserted, we can apply a local unitary transformation to ``integrate'' the layer into the bulk. The anyons that come from the layers can acquire a different (but still local) profile in the $z$ direction when becoming a planon. In particular, if the unitaries have exponentially decaying tails, the profile can have exponential tails, which is not surprising. But it is not possible for the planons to have exponentially decaying tails in its statistics because unitary transformations cannot change statistics. The only thing we can do when mapping the anyons in the foliation layers into planons is to relabel them, i.e. choose a different set and call them elementary. But when combining anyons into a new generating set, it is not possible to combine fractions of them together in the form of an exponentially decaying tail. Therefore, the exponential decaying infinite statistics precludes a foliation structure. Moreover, this ``profile" of braiding statistics defines an intrinsic length scale in the system along the $z$ direction, determined entirely by the topological order and can not be tuned continuously. As we show below, the length scale characterizes the spread of anyon string operators in the $z$ direction.
 
Similar phenomena can be found in many other iCS theories, as discussed in Appendix~\ref{sec:det}. In fact, the properties of $K_{\text{nF}}$ are so unusual that one may wonder if it represents a physical $3+1$D theory at all and if so, whether the planons are indeed point excitations. In Ref.~\onlinecite{Qiu90,Qiu90-2, Naud00, Naud01}, the theory was studied in terms of its related Laughlin wave-function and the corresponding quantum Hall Hamiltonian, which partially addresses this question. In section~\ref{sec:lattice}, we address this question for all iCS theories with quasi-diagonal $K$ matrices through explicit lattice construction. We show that all such theories are local $3+1$D models and in particular for $K_{\text{nF}}$, the elementary planons are indeed point excitations. They move in the $xy$ plane and are hence planons.

\section{Gapless theories}
\label{sec:gapless}

If we change the diagonal entries in Eq.~\ref{eq:KnF} from $3$ to $2$, 
\begin{equation}
    \label{eq:K_gl}
    K_{\text{gl}}=\begin{pmatrix}
    2 & 1 & & & 1 \\
    1 & 2 & 1 & & \\
    & \ddots & \ddots & \ddots & \\
    & & 1 & 2 & 1 \\
    1 & & & 1 & 2
    \end{pmatrix},
\end{equation}
we get a very different theory. In particular, the calculation in section~\ref{sec:spectrum} shows that the theory becomes gapless. The ``gl'' subscript denotes ``gapless''. It is not clear what the nature of the gapless phase is. In this section, we will simply list some of the properties of $K_{\text{gl}}$.

The eigenvalues of $K_{\text{gl}}$ form a gapless band with a quadratic dispersion. Therefore, according to discussion in section~\ref{sec:spectrum}, the photon sector in the theory is gapless with a quadratic dispersion in the $z$ direction. As the band touches the zero energy point when the size $N$ of the matrix is even, the determinant of the matrix is zero with even $N$. With odd $N$, the determinant is always 4. 

The inverse of the matrix looks like, for $N=5$,
\begin{equation*}
    K_{\text{gl}}^{-1} = \frac{1}{4} \begin{pmatrix} 5 & -3 & 1 & 1 & -3 \\ -3 & 5 & -3 & 1 & 1 \\ 1 & -3 & 5 & -3 & 1\\ 1 & 1 & -3 & 5 & -3 \\ -3 & 1 & 1 & -3 & 5 \end{pmatrix},
\end{equation*}
while for $N=7$,
\begin{equation*}
K_{\text{gl}}^{-1} = \frac{1}{4} \begin{pmatrix} 7 & -5 & 3 & -1 & -1 & 3 & -5 \\ -5 & 7 & -5 & 3 & -1 & -1 & 3 \\ 3 & -5 & 7 & -5 & 3 & -1 & -1\\ -1 & 3 & -5 & 7 & -5 & 3 & -1 \\ -1 & -1 & 3 & -5 & 7 & -5 & 3 \\ 3 & -1 & -1 & 3 & -5 & 7 & -5 \\ -5 & 3 & -1 & -1 & 3 & -5 & 7\end{pmatrix}.
\end{equation*}
The fusion group in this case turns out to be $\mathbb{Z}_4$ and the topological spin of the generating anyon is $\theta=q\pi/4$, where $q=N\text{ mod }8$. Hence, the topological order is that of the $\nu=2q$ fermionic $\mathbb{Z}_2$ gauge theory in Kitaev's 16-fold way.\cite{KitaevAnyons} The entries in $K_{\text{gl}}^{-1}$ decay linearly away from the main diagonal. However, unlike for $K_{\text{nF}}$ in Eq.~\ref{eq:KnF}, the statistics do not become more fractional as the system size grows. Instead, the fractional part remains $\pm 1/4$ no matter the distance. Because of this, the fractional excitations are hence very different from those in $K_{\text{nF}}$. As we will show in section~\ref{sec:lattice}, while the fractional excitations in $K_{\text{nF}}$ have a localized profile in the $z$ direction and can be considered as point excitations, those in $K_{\text{gl}}$ have an extensive profile in the $z$ direction and should be regarded as a line excitation (if such consideration is valid at all given the existence of gapless modes in the model).

$K_{\text{gl}}$ is a representative of the class of gapless iCS theories with quasi-diagonal $K$ matrices. In Ref.~\onlinecite{Qiu90-2}, it was mentioned that some of these theories might have an instability towards ``staging'', that is, translation symmetry breaking in the $z$ direction. Whether that always happens, or whether some of these theories might be gapless spin liquids or gapless fracton phases is not clear. We will leave more in-depth study of these gapless phases to future work. 

\section{Lattice Construction}
\label{sec:lattice}

In the previous sections, we have presented some interesting and sometimes even surprising properties of the iCS theories without addressing one crucial question: are the iCS theories legitimate $3+1$D models? In particular, can we interpret the $i$ index in Eq.~\ref{eq:lagrangian} as a $z$ direction spatial coordinate? After all, the Chern-Simons gauge fields $\mathcal{A}^i$ are not local degrees of freedom and can have complicated commutation relations between one another. For example, when $e\to \infty$, 
\begin{equation*}
    \left[\mathcal{A}^i_x,\mathcal{A}^j_{y}\right] \propto K^{-1}_{ij}.
\end{equation*}
The situation is particularly worrisome in the case of $K_{\text{nF}}$ and $K_{\text{gl}}$ where the entries in $K^{-1}$ are all nonzero. It means that if we try to interpret $i$ and $j$ as the $z$ direction spatial coordinate, the gauge field in the $i$th layer $\mathcal{A}^i$ would have nontrivial commutation relation with the gauge field in the $j$th layer even though they are very far away.

This is related to the question of what the fractional excitations look like, in particular whether the ones associated with the unit vectors $(\ldots,0,0,1,0,0,\ldots)$ have a local profile in the $z$ direction. In the CS formulation, this seems to be the case because these excitations are unit gauge charges of the gauge field $\mathcal{A}^i$ and are created simply by string operators of the form (in the $e\to \infty$ limit)
\begin{equation*}
    \mathcal{W}^i=\exp\left[-i\int_{\text{path}} dx_\alpha\mathcal{A}^i_\alpha\right],\label{eq:string_field}
\end{equation*}
but this seems to be at odds with the fact that the $i$th excitation has a nontrivial braiding statistic with the $j$th excitation no matter how far away they are.

In this section, we clarify these issues by presenting a lattice construction whose low energy effective theory is described by Eq.~\ref{eq:lagrangian}. Our construction works for any iCS theory with a quasi-diagonal $K$ matrix --- symmetric integer matrices whose entries are zero beyond a certain distance from the main diagonal and whose nonzero entries are all bounded by some finite number. Therefore, our construction shows that all such iCS theories are legitimate $3+1$D local models. Moreover, we write down the explicit form of the string operators that generate fractional excitations and show that for $K_{\text{F}}$, $K_{\text{nF}}$, the elementary excitations --- those associated with unit vectors $(\ldots,0,0,1,0,0,\ldots)$ --- are local in the $z$ direction and are hence point excitations. For $K_{\text{gl}}$, however, the elementary excitation is not localized in the $z$ direction and should not be thought of as a point excitation.

\subsection{Lattice model}
\label{subsec:model}

We now describe the lattice model that realizes a quasi-diagonal iCS theory. For clarity, we start with a toy example $K=(2)$, a $1\times1$ matrix. Although this $K$ matrix has finite dimension, it contains much of the relevant physics, and will also be revisited in Section~\ref{subsec:string} when we study string operators. We then proceed to the less trivial example of $K_{\text{nF}}$ defined in Eq.~\ref{eq:KnF}. Finally, we present the construction in full generality which works for arbitrary quasi-diagonal $K$ with bounded entries.

\begin{figure}[ht]
    \centering
    \includegraphics[width=0.231\textwidth]{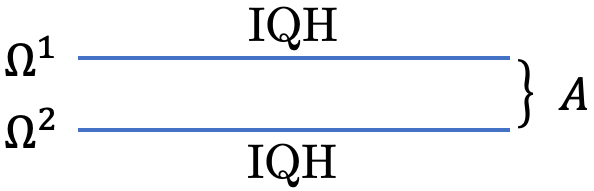}
    \caption{Lattice model realizing $K=(2)$. The matter content of the system is two IQH layers $\Omega^1$ and $\Omega^2$ (blue lines) with Chern number $C^l=1$. The layers are coupled each with unit charge to a dynamical $U(1)$ gauge fields $A$. }
    \label{fig:eg_1}
\end{figure}

\begin{figure}[t]
    \centering
    \includegraphics[width=0.471\textwidth]{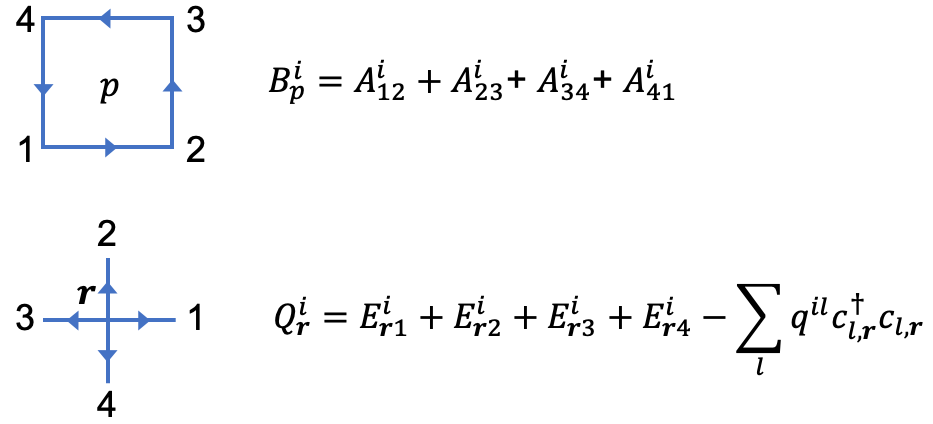}
    \caption{The flux and Gauss's law terms in Eq.~\ref{eq:hamiltonian}. Reversing the direction of the edges changes the signs in front of $A$ and $E$. In the context of our first example $K=(2)$, the index $i$ should be ignored and $q^{il}=1$ for $l=1,2$.}
    \label{fig:B_and_Q}
\end{figure}

The $K=(2)$ CS theory can be realized as a chiral spin liquid, as discussed for example in Ref.~\onlinecite{Kalmeyer87}. 
Here we present a more complicated construction so that it can be generalized to all iCS theories. We start with two integer quantum Hall (IQH) layers $\Omega^l$, $l=1,2$, with Chern number $C^l=1$. Each layer is a free fermion hopping model in the $xy$ plane. The fermions in each layer carry unit charge under a global charge conservation symmetry and we can gauge the system by coupling the layer to a dynamical $U(1)$ gauge field $A$. More precisely, we add gauge degrees of freedom $A_{\mathbf{r}\mathbf{r}'}$ on the horizontal links $\left<\mathbf{r}\mathbf{r}'\right>$. As usual, we define the electric field $E_{\mathbf{r}\mathbf{r}'}$ as the conjugate variable to $A_{\mathbf{r}\mathbf{r}'}$, $[A_{\mathbf{r}\mathbf{r}'},E_{\mathbf{r}\mathbf{r}'}]=i$. The Hamiltonian after gauging is
\begin{align}
\label{eq:toy_hamiltonian}
    H=&\sum_{l=1,2}\sum_{\left<\mathbf{r}\mathbf{r}'\right>}u_{\mathbf{r}\mathbf{r}'}e^{iA_{\mathbf{r}\mathbf{r}'}}c^{\dagger}_{l,\mathbf{r}'}c_{l,\mathbf{r}}+\sum_{\left<\mathbf{r}\mathbf{r}'\right>}g_E\left(E_{\mathbf{r}\mathbf{r}'}\right)^2\nonumber\\
    &-g_B\sum_p\cos B_p+g_Q\sum_\mathbf{r}\left(Q_\mathbf{r}\right)^2,
\end{align}
where $\mathbf{r}$, $\mathbf{r}'$ are 2-component vectors labelling the sites in each layer, $u_{\mathbf{r}\mathbf{r}'}$ is the IQH hopping coefficient, $B_p$ is the flux of $A$ through plaquette $p$, and
\begin{equation}
\label{eq:toy_gauss}
    Q_\mathbf{r}=(\nabla\cdot\mathbf{E})_\mathbf{r} -\sum_{l=1,2}c_{l,\mathbf{r}}^{\dagger}c_{l,\mathbf{r}}
\end{equation}
is the Gauss's law term (see Fig.~\ref{fig:B_and_Q}). Note that here Gauss's law is only being imposed as an energetic constraint, not a Hilbert space constraint. Because of this, the resulting theory is fermionic instead of bosonic. More specifically, we will show below that the resulting theory is the $K=2$ bosonic theory together with two decoupled fermionic IQH layers.

At low energies, the model is described by an effective CS theory (we kept only the topological CS terms and omitted the Maxwell term and source term $A_\mu J^\mu$)
\begin{equation}
    \mathcal{L}=-\frac{1}{4\pi}\sum_{l=1,2}C^l \epsilon^{\mu\nu\lambda} a^l_{\mu}\partial_{\nu}a^l_{\lambda}+\frac{1}{2\pi}\sum_{l=1,2}\epsilon^{\mu\nu\lambda}A_\mu\partial_\nu a_\lambda^l,
    \label{eq:toy_L}
\end{equation}
whose $K$ matrix with respect to the basis $(a^1,a^2,A)$ is
\begin{equation}
\label{eq:toy_K0}
    K_0=
    \begin{pmatrix}
    -1 & 0 & 1\\
    0 & -1 & 1\\
    1 & 1 & 0
    \end{pmatrix}.
\end{equation}
Note that an IQH layer with Chern number 1 corresponds to a $-1$ in the $K$ matrix. To see how $K_0$ relates to the desired $K=(2)$, we apply the transformation $K_0\mapsto\widetilde{K}_0=WK_0W^T$ with
\begin{equation*}
    W=
    \begin{pmatrix}
    1 & 0 & 0\\
    0 & 1 & 0\\
    1 & 1 & 1
    \end{pmatrix}.
\end{equation*}
We obtain
\begin{equation*}
    \widetilde{K}_0=
    \begin{pmatrix}
    -1 & 0 & 0\\
    0 & -1 & 0\\
    0 & 0 & 2
    \end{pmatrix}
\end{equation*}
in terms of the new fields
\begin{equation*}
    \begin{pmatrix} \tilde{a}^1 \\ \tilde{a}^2 \\ \widetilde{A} \end{pmatrix} = \left(W^{-1}\right)^T\begin{pmatrix} a^1 \\ a^2 \\ A \end{pmatrix} = \begin{pmatrix} a^1-A \\ a^2 -A \\ A \end{pmatrix}.
\end{equation*}
We see that $\widetilde{K}_0$ contains the decoupled block $K=(2)$ in its lower right corner. We also have two decoupled IQH layers in $\widetilde{K}_0$, but these have no anyon content. Therefore, the construction, as written, realizes not exactly the $K=2$ theory, but a very close fermionic cousin represented by $\tilde{K}_0$.

\begin{figure}[ht]
    \centering
    \includegraphics[width=0.233\textwidth]{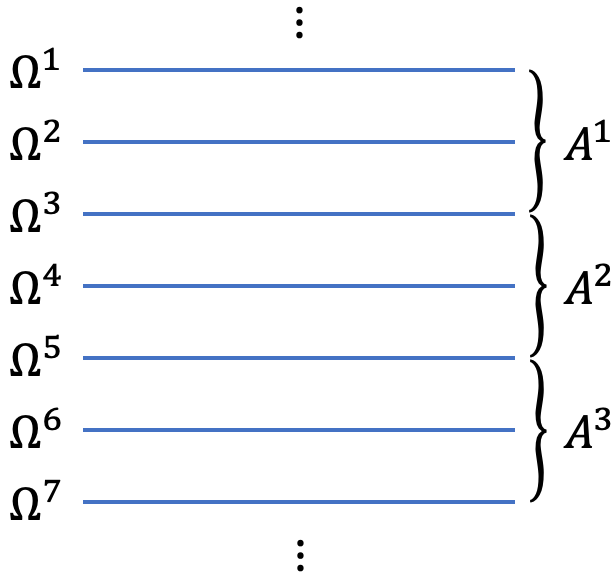}
    \caption{Lattice model realizing $K_{\text{nF}}$. The matter content of the system is infinitely many IQH layers $\Omega^l$ (blue lines) with Chern number $C^l=1$. The layers are coupled with unit charge to infinitely many dynamical $U(1)$ gauge fields $A^i$ in the way indicated by the curly brackets. }
    \label{fig:eg_2}
\end{figure}

Next, we consider the example of $K_{\text{nF}}$ defined in Eq.~\ref{eq:KnF}. To realize $K_{\text{nF}}$, we take infinitely many IQH layers $\Omega^l$, $l\in\mathbb{Z}$, each with Chern number $C^l=1$. We couple the layers to infinitely many dynamical $U(1)$ gauge fields $A^i$, $i\in\mathbb{Z}$, as follows: fermions in layers $\Omega^1$, $\Omega^2$, $\Omega^3$ have unit charge under $A^1$, those in layers $\Omega^3$, $\Omega^4$, $\Omega^5$ have unit charge under $A^2$, those in layers $\Omega^5$, $\Omega^6$, $\Omega^7$ have unit charge under $A^3$, etc. All other pairs of $\Omega^l$ and $A^i$ not following this pattern are uncoupled (see Fig.~\ref{fig:eg_2}). This model has a low energy effective CS theory similar to Eq.~\ref{eq:toy_L}, but now with $K$ matrix
\begin{equation*}
    K_0=
    \begin{pmatrix}
    \ddots\\
    & -1 & & 1\\
    & & -1 & 1\\
    & 1 & 1 & 0 & 1\\
    & & & 1 & -1 & & 1\\
    & & & & & -1 & 1\\
    & & & & 1 & 1 & 0 & 1\\
    & & & & & & 1 & -1\\
    & & & & & & & & \ddots
    \end{pmatrix}.
\end{equation*}
with respect to the basis $\left(...,a^1,a^2,A^1,a^3,a^4,A^2,a^5,...\right)$. Like in the previous example, we apply the transformation $K_0\mapsto\widetilde{K}_0=WK_0W^T$ with
\begin{equation*}
    W=
    \begin{pmatrix}
    \ddots\\
    & 1\\
    & & 1\\
    & 1 & 1 & 1 & 1\\
    & & & & 1\\
    & & & & & 1\\
    & & & & 1 & 1 & 1 & 1\\
    & & & & & & & 1\\
    & & & & & & & & \ddots
    \end{pmatrix}.
\end{equation*}
$W$ is a local transformation in the sense that it can be decomposed into two layers where each layer is a product of non-overlapping general linear transformations that act on three nearest neighbor dimensions. Borrowing the terminology for local unitary transformations, $W$ is a ``finite depth circuit'' of general linear transformations. This is an important point because it shows that locality is preserved when we map from the lattice model to the effective iCS theory. Specifically, $W$ can be decomposed into a finite product of block diagonal integer matrices $W=W_1W_2$, where
\begin{equation*}
    W_1=
    \begin{pmatrix}
    \ddots\\
    & 1\\
    & & 1\\
    & 1 & & 1\\
    & & & & 1\\
    & & & & & 1\\
    & & & & 1 & & 1\\
    & & & & & & & 1\\
    & & & & & & & & \ddots
    \end{pmatrix},
\end{equation*}
\begin{equation*}
    W_2=
    \begin{pmatrix}
    \ddots\\
    & 1\\
    & & 1\\
    & & 1 & 1 & 1\\
    & & & & 1\\
    & & & & & 1\\
    & & & & & 1 & 1 & 1\\
    & & & & & & & 1\\
    & & & & & & & & \ddots
    \end{pmatrix}.
\end{equation*}
The result of this transformation is
\begin{equation*}
    \widetilde{K}_0=
    \begin{pmatrix}
    \ddots\\
    & -1\\
    & & -1\\
    & & & 3 & & & 1\\
    & & & & -1\\
    & & & & & -1\\
    & & & 1 & & & 3\\
    & & & & & & & -1\\
    & & & & & & & & \ddots
    \end{pmatrix},
\end{equation*}
which breaks into $K_{\text{nF}}$ consisting of rows and columns with indices $...,3,6,9,...$ together with decoupled $\nu=1$ IQH layers. Similar to the previous case, we get almost the theory we want except for some extra IQH layers which do not have any impact on the anyon statistics of the theory.

Having discussed two examples, we finally present a construction that works for an arbitrary quasi-diagonal $K$ with bounded entries. Similar to the $K_{\text{nF}}$ example, we start with a stack of IQH layers $\Omega^l$. The Chern number of layer $l$ is $C^l=\pm1$, to be fixed later. We introduce gauge degrees of freedom $A_{\mathbf{r}\mathbf{r}'}^i$ and their conjugate momenta $E_{\mathbf{r}\mathbf{r}'}^i$ on the horizontal links $\left<\mathbf{r}\mathbf{r}'\right>$ and impose the commutation relation $[A_{\mathbf{r}\mathbf{r}'}^j,E_{\mathbf{r}\mathbf{r}'}^k]=i\delta_{jk}$ as usual. We then couple $\Omega^l$ to $A^i$ with charge $q^{il}$, also to be fixed later. The resulting Hamiltonian is
\begin{multline}
\label{eq:hamiltonian}
    H=\sum_l\sum_{\left<\mathbf{r}\mathbf{r}'\right>}u_{l,\mathbf{r}\mathbf{r}'}\exp{\left(i\sum_i q^{il}A^i_{\mathbf{r}\mathbf{r}'}\right)}c^{\dagger}_{l,\mathbf{r}'}c_{l,\mathbf{r}}\\
    +\sum_i\left[\sum_{\left<\mathbf{r}\mathbf{r}'\right>}g_E\left(E_{\mathbf{r}\mathbf{r}'}^i\right)^2-g_B\sum_p\cos B_p^i\right.\\
    \left.+g_Q\sum_\mathbf{r}\left(Q_\mathbf{r}^i\right)^2\right],
\end{multline}
where $u_{l,\mathbf{r}\mathbf{r}'}$ is the IQH hopping coefficient determined by $C^l$, $B_p^i$ is the flux of $A^i$ through plaquette $p$, and
\begin{equation}
\label{eq:lattice_gauss}
    Q_\mathbf{r}^i=(\nabla\cdot\mathbf{E})_\mathbf{r}^i -\sum_l q^{il} c_{l,\mathbf{r}}^{\dagger}c_{l,\mathbf{r}}
\end{equation}
is the Gauss's law term (see Fig.~\ref{fig:B_and_Q}). We think of the fermion and gauge field layers as interlaced in the $z$ direction. The interactions are local as long as only a finite number of neighboring layers are charged under each $A^i$, or equivalently, each row and column of $q^{il}$ has bounded support, which turns out to hold with our choice of $q^{il}$ later. The low energy field theory of Eq.~\ref{eq:hamiltonian} is given by
\begin{equation}
\label{eq:general_L}
    \mathcal{L}=-\frac{1}{4\pi}\sum_{l}C^l \epsilon^{\mu\nu\lambda} a^l_{\mu}\partial_{\nu}a^l_{\lambda}+\frac{1}{2\pi}\sum_{il}q^{il}\epsilon^{\mu\nu\lambda}A_\mu^i\partial_\nu a_\lambda^l.
\end{equation}
Here we have only kept the CS terms and omitted the Maxwell and source terms.

To realize a particular $K=(K_{ij})$, we need to specify $\Omega^l$, $C^l$ and $q^{il}$. We adopt the following setup:
\begin{enumerate}[leftmargin=*]
    \item For each index $i$ of $K$, we have a dynamical $U(1)$ gauge field $A^i$.
    
    \item For each $i$ such that \[\Delta_i:=K_{ii}-\sum_{j\neq i}K_{ij}\neq0,\] we have IQH layers $\Omega^{i,s}_\text{d}$ where $s=1,2,...,|\Delta_i|$ and the subscript ``d'' stands for ``diagonal''. Each $\Omega^{i,s}_\text{d}$ has Chern number $C_\text{d}^i=\text{sgn}(\Delta_i)$ and carries $+1$ charge under $A^i$ only. The emergent gauge field of $\Omega^{i,s}_\text{d}$ is denoted by $a_\text{d}^{i,s}$. If $\Delta_i=0$, no diagonal layer is needed.
    
    \item For each pair $i<j$ such that $K_{ij}\neq0$, we have IQH layers $\Omega^{ij,t}_\text{o}$ where $t=1,2,...,|K_{ij}|$ and the subscript ``o'' stands for ``off-diagonal''. Each $\Omega^{ij,t}_\text{o}$ has Chern number $C_\text{o}^{ij}=\text{sgn}(K_{ij})$ and carries $+1$ charge under $A^i$ and $A^j$ only. The emergent gauge field of $\Omega^{ij,t}_\text{o}$ is denoted by $a_\text{o}^{ij,t}$.
\end{enumerate}
Since $K$ is quasi-diagonal with bounded entries, all these IQH layers $\Omega$ and physical gauge fields $A$ can be interlaced in the $z$-direction in such a way that the interaction is local. We denote by $\mathbfcal{A}$ the collection of emergent and physical gauge fields ordered in this particular way, and $K_0$ the $K$ matrix of the CS theory Eq.~\ref{eq:general_L} with respect to the basis $\mathbfcal{A}$. Next, we apply the local transformation $\widetilde{\mathcal{A}}^i=\sum_j \left(W^{-1}\right)^{ji}\mathcal{A}^j$, $\widetilde{K}_0=WK_0W^T$ defined by
\begin{align*}
    \tilde{a}_\text{d}^{i,s}&=a_\text{d}^{i,s}-\text{sgn}(\Delta_i)A^i,\\
    \tilde{a}_\text{o}^{ij,t}&=a_\text{o}^{ij,t}-\text{sgn}(K_{ij})\left(A^i+A^j\right),\\
    \widetilde{A}^i&=A^i.
\end{align*}
This transformation is local in the sense that $W$ can be decomposed into a finite depth circuit (i.e. a finite product) of local, block diagonal integer matrices. In fact, the circuit has depth 2. The first step of the circuit is to map
\begin{align*}
    a_\text{d}^{i,s}&\mapsto a_\text{d}^{i,s}-\text{sgn}(\Delta_i)A^i,\\
    a_\text{o}^{ij,t}&\mapsto a_\text{o}^{ij,t}-\text{sgn}(K_{ij})A^i,
\end{align*}
and the second step is to map \[a_\text{o}^{ij,t}\mapsto a_\text{o}^{ij,t}-\text{sgn}(K_{ij})A^j.\] Each step is block diagonal because each $a$ is modified by at most one $A$, and each block is local because we have arranged the degrees of freedom in the $z$ direction such that each $A^i$ is some finite distance away from each $a_\text{d}^{i,s}$ and $a_\text{o}^{ij,t}$. After the transformation, the $\tilde{a}_\text{d}$ and $\tilde{a}_\text{o}$ fields are in decoupled IQH states, and the $\widetilde{A}$ fields have the desired $K$ matrix.

We conclude this subsection by relating the general construction to the two examples we gave. For $K=(2)$, we have
\begin{align*}
    \mathbfcal{A}&=\left(a^1,a^2,A\right)\nonumber\\
    &=\left(a_\text{d}^{1,1},a_\text{d}^{1,1},A^1\right),
\end{align*}
with no ``off-diagonal'' layers. For $K_{\text{nF}}$, we have
\begin{align*}
    \mathbfcal{A}&=\left(...,a^1,a^2,A^1,a^3,a^4,A^2,a^5,...\right)\nonumber\\
    &=\left(...,a_\text{o}^{01,1},a_\text{d}^{1,1},A^1,a_\text{o}^{12,1},a_\text{d}^{2,1},A^2,a_\text{o}^{23,1},...\right).
\end{align*}

\subsection{Spectrum of iCS theory}
\label{sec:spectrum}

Given an iCS theory, we can calculate its spectrum from its Lagrangian Eq.~\ref{eq:lagrangian}. Note that it is important to include the Maxwell term for this calculation. In the temporal gauge $A_0=0$, the equations of motion are
\begin{align*}
    \partial_t^2A_x^i+\partial_x\partial_yA_y^i-\partial_y^2A_x^i+\frac{e^2}{2\pi}K_{ij}\partial_tA_y^j&=0,\\
    \partial_t^2A_y^i+\partial_x\partial_yA_x^i-\partial_x^2A_y^i-\frac{e^2}{2\pi}K_{ij}\partial_tA_x^j&=0.
\end{align*}
They are solved by \[A_{x,y}^i=\alpha_{x,y}v^i_q\exp{\left[i(k_xx+k_yy-\omega t)\right],}\]
where $v^i_q$ is an eigenvector of $K$ with eigenvalue $K_q$, labelled by $q$. We find the spectrum \[\omega^2=k_x^2+k_y^2+\left(\frac{e^2}{2\pi}K_q\right)^2.\] If $K$ is invariant under translation along the diagonal such as $K_{\text{nF}}$ and $K_{\text{gl}}$, then $q$ is the momentum in the $z$ direction. For $K_{\text{nF}}$ we have $K_q=3+2\cos q$ therefore the whole spectrum is gapped. For $K_{\text{gl}}$ we have $K_q=2+2\cos q$ which is gapless and the full spectrum has a zero mode at momentum $(k_x,k_y,q) = (0,0,\pi)$.

\subsection{String operators}
\label{subsec:string}

We now study the string operators of the fractional excitations in our lattice model Eq.~\ref{eq:hamiltonian}. We work in the limit of $g_E=0$, and will argue later about the case where $g_E$ is nonzero but small. For simplicity, we first consider the example $K=(2)$ studied in section~\ref{subsec:model}, for which we wrote down a lattice model with low energy CS theory given by Eq.~\ref{eq:toy_K0}. This system contains one type of fractional excitation, which is a semion. A charge vector that lies in the semion superselection sector is $Q=(0,0,1)^T$; the general form of a semion charge vector is $(-a,-b,a+b+2c+1)^T$ where $a,b,c\in\mathbb{Z}$. The flux vector attached to $Q$ is
\begin{equation}
\label{eq:toy_attach}
    \Phi=-2\pi K_0^{-1}Q=
    \begin{pmatrix}
    -\pi \\ -\pi \\ -\pi
    \end{pmatrix}.
\end{equation}
The $-\pi$ fluxes for the emergent fields $a^1$ and $a^2$ should be interpreted as $-1/2$ fermion charges in each fermion layer. Therefore, the semion consists of a $+1$ \textit{external} charge, a $-\pi$ dynamical flux, a $-1/2$ charge in $\Omega^1$ and a $-1/2$ charge in $\Omega^2$.

\begin{figure}[t]
    \centering
    \includegraphics[width=0.48\textwidth]{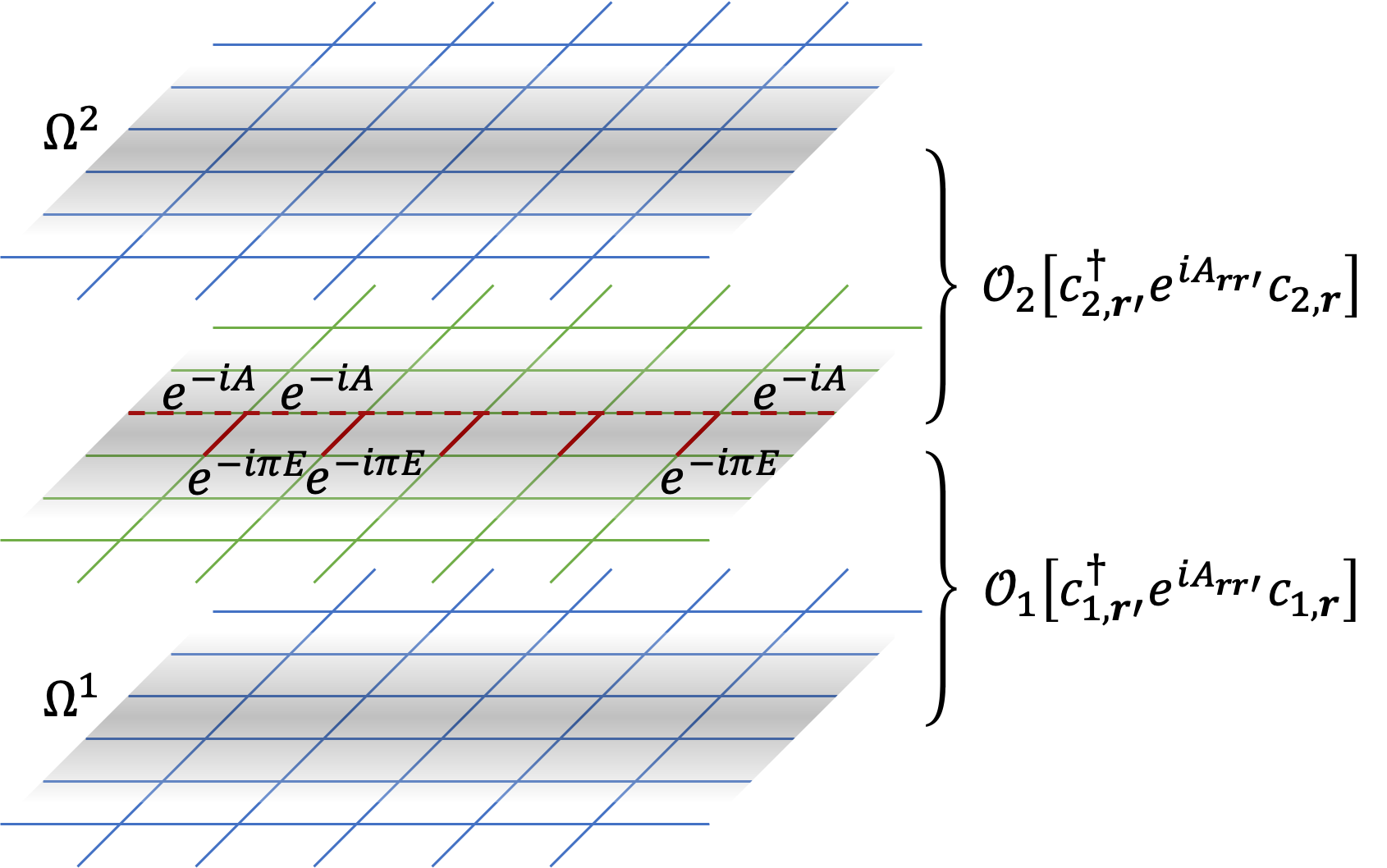}
    \caption{The string operator for the lattice model realizing $K=(2)$. Fermions live in the blue layers $\Omega^1$ and $\Omega^2$, and the gauge field in the middle, green layer. The operators $\mathcal{O}_l$ are generated by hopping operators $c^{\dagger}_{l,\mathbf{r}'}e^{iA_{\mathbf{r}\mathbf{r}'}}c_{l,\mathbf{r}}$. The action of $\mathcal{O}_l$ is non-trivial only near the path (grey region), and is exponentially close to the identity away from the path. The string operator $\mathcal{W}$ consists of $e^{-iA}$ acting on the dashed red line, $e^{-i\pi E}$ acting on the solid red segments and $\mathcal{O}_1$, $\mathcal{O}_2$ acting near the path.}
    \label{fig:string}
\end{figure}

The string operator $\mathcal{W}$ consists of three parts, $\mathcal{W}=\mathcal{W}_1\mathcal{W}_2\mathcal{W}_3$, as follows (see Fig.~\ref{fig:string}):
\begin{enumerate}[leftmargin=*]
    \item $\mathcal{W}_1=\prod_{\text{path}}e^{-iA}$ acts on the dynamical gauge field along the path and creates a $+1$ external charge at the end of the path (and a $-1$ charge at the start).
    
    \item $\mathcal{W}_2=\prod_{\perp\text{path}} e^{-i\pi E}$ acts on the dynamical gauge field along adjacent links to the right of and perpendicular to the path, and creates a $-\pi$ flux at the end of the path. $\mathcal{W}_2$ acts only on the gauge DOF.
    
    \item $\mathcal{W}_3$ is the quasi-adiabatic response\cite{Hastings2005} of the fermions to the $-\pi$ flux insertion. More precisely, in each gauge field sector $\{A_{\mathbf{r}\mathbf{r}'}\}$ of the Hilbert space, we insert an external $-\pi$ flux adiabatically, which is implemented by an $A$-dependent evolution operator $\mathcal{W}_3[A]$ on the fermion Hilbert space. As the fermion hopping model is not exactly solvable, we do not know the exact expression for $\mathcal{W}_3[A]$ except that it is of the form \[\mathcal{W}_3[A]=\mathcal{O}_1\left[c^{\dagger}_{1,\mathbf{r}'}e^{iA_{\mathbf{r}\mathbf{r}'}}c_{1,\mathbf{r}}\right] \mathcal{O}_2\left[c^{\dagger}_{2,\mathbf{r}'}e^{iA_{\mathbf{r}\mathbf{r}'}}c_{2,\mathbf{r}}\right],\] where $\mathcal{O}_l\left[c^{\dagger}_{l,\mathbf{r}'}e^{iA_{\mathbf{r}\mathbf{r}'}}c_{l,\mathbf{r}}\right]$ is some gauge invariant operator generated by hopping operators $c^{\dagger}_{l,\mathbf{r}'}e^{iA_{\mathbf{r}\mathbf{r}'}}c_{l,\mathbf{r}}$. Nonetheless, properties of quasi-adiabatic evolution ensure that $\mathcal{W}_3[A]$ is local, acting only near the path (grey region in Fig.~\ref{fig:string}). A $-1/2$ charge in $\Omega^1$ and a $-1/2$ charge in $\Omega^2$ are accumulated in the process near the end of the string operator, which correspond to the $-\pi$ fluxes of $a^1$ and $a^2$.
\end{enumerate}

We check the correctness of our string operator by computing the semion braiding phase, which we expect to be $2\pi Q^T K_0^{-1}Q=\pi$. To see this from the string operator, we break the overall commutation relation into the commutations of:
\begin{enumerate}[leftmargin=*]
    \item $\mathcal{W}_1$ with $\mathcal{W}_2$. This takes a $+1$ charge counterclockwise around a $-\pi$ flux, giving a phase of $\pi$.
    
    \item $\mathcal{W}_2$ with $\mathcal{W}_1$. This gives a phase of $\pi$ for the same reason.
    
    \item the product $\mathcal{W}_2\mathcal{W}_3$ with itself. This contributes a phase $-\pi$ which can be understood as the Berry phase obtained due to the following actions on the fermions: increasing the (background) flux in the $x$ direction by $\pi$, increasing the flux in the $y$ direction by $\pi$, decreasing the flux in the $x$ direction by $\pi$, decreasing the flux in the $y$ direction by $\pi$. In each IQH layer, the Berry phase over the entire flux parameter space $[0,2\pi)^2$ is $-2\pi$. The Berry phase over a quarter of the parameter space is therefore $-\pi/2$. As we have two IQH layers, the total phase is $-\pi$.
    
    \item $\mathcal{W}_1$ with itself, $\mathcal{W}_1$ with $\mathcal{W}_3$ and $\mathcal{W}_3$ with $\mathcal{W}_1$. All of these are trivial.
\end{enumerate}
Summing these contributions up, we find a total braiding phase $\pi+\pi-\pi=\pi$ as expected. Of course, phases are defined mod $2\pi$, but we have been careful distinguishing e.g.~$-\pi$ from $\pi$ so that this argument extends naturally to general $K$.

So far we have considered the $g_E=0$ limit, where we showed that $\mathcal{W}$ is a string operator for the charge vector $Q=(0,0,1)^T$. In fact, in this limit we could write down many other different string operators for $Q$ which all commute with the Hamiltonian except near the end points. For example, we could have $\mathcal{W}'=\mathcal{W}'_1\mathcal{W}'_2\mathcal{W}'_3$ where $\mathcal{W}'_1=\prod_{\text{path}}e^{-iA}$ as before, $\mathcal{W}'_2=\prod_{\perp\text{path}}e^{i\theta E}$ for arbitrary $\theta$ and $\mathcal{W}'_3$ is the quasi-adiabatic response of the fermions to a $\theta$ flux insertion. To see why we chose the particular $\mathcal{W}$ that satisfies the charge-flux attachment condition Eq.~\ref{eq:toy_attach}, we turn on a small $g_E>0$, much smaller than the other couplings in the Hamiltonian and the Landau level spacing. Now if the string operator creates a $\theta$ flux and hence a $\theta/2\pi$ charge in each IQH layer, then Gauss's law (Eq.~\ref{eq:toy_gauss}) implies \[\nabla\cdot\mathbf{E}=1+\frac{\theta}{\pi}.\] If $\nabla\cdot\mathbf{E}\neq 0$, then we have an electric energy that diverges at least logarithmically. Therefore, we must choose $\theta=-\pi$ so that $\nabla\cdot\mathbf{E}=0$. This way, when $g_E>0$, it is possible to modify $\mathcal{W}$ in a region near the path such that the electric energy is finite. Furthermore, since $g_E$ is small, the gauge field sectors $\{A_{\mathbf{r}\mathbf{r}'}\}$ that are present in the ground state can differ from the flat configuration $B\equiv0$ at most by a small perturbation. Therefore, even with the new hopping coefficients $u_{l,\mathbf{r}\mathbf{r}'}e^{iA_{\mathbf{r}\mathbf{r}'}}$, the fermions are still in a $C^l=1$ IQH state, so the $-\pi$ flux is indeed bound with a $-1/2$ charge in each layer. The exact expression of the new $\mathcal{W}$ is not important, and the braiding statistic remains unchanged as long as the correct amount of external charge, fermion charge and flux are created.

A similar construction of string operators works for iCS theories. When $g_E=0$, the string operator $\mathcal{W}^i$ corresponding to standard basis vector $e_i$ takes the form $\mathcal{W}^i=\mathcal{W}^i_1\mathcal{W}^i_2\mathcal{W}^i_3$. First, $\mathcal{W}_1^i=\prod_{\text{path}}e^{-iA^i}$ creates a $+1$ external $A^i$ charge. Next,
\begin{equation*}
    \mathcal{W}^i_2=\prod_{\perp\text{path}}\exp\left[-2\pi i\sum_j\left(K^{-1}\right)^{ij}E^j\right]
\end{equation*}
creates fluxes according to the $i$th row of $K^{-1}$, as required by Gauss's law Eq.~\ref{eq:lattice_gauss} when a small $g_E>0$ is present. The IQH layers then respond quasi-adiabatically, giving an evolution operator $\mathcal{W}^i_3$. The braiding statistic of $\mathcal{W}^i$ and $\mathcal{W}^j$ results from the commutations of $\mathcal{W}^i_1$ with $\mathcal{W}^j_2$, $\mathcal{W}^i_2$ with $\mathcal{W}^j_1$, and $\mathcal{W}^i_2\mathcal{W}^i_3$ with $\mathcal{W}^j_2\mathcal{W}^j_3$. In particular, the commutation of $\mathcal{W}^i_2\mathcal{W}^i_3$ with $\mathcal{W}^j_2\mathcal{W}^j_3$ correspond to the following actions on the fermions: increasing the (background) $A^k$ flux in the $x$ direction by $2\pi\left(K^{-1}\right)^{ik}$ for all $k$, increasing the $A^l$ flux in the $y$ direction by $2\pi\left(K^{-1}\right)^{jl}$ for all $l$, decreasing the $A^k$ flux in the $x$ direction by $2\pi\left(K^{-1}\right)^{ik}$ for all $k$, decreasing the $A^l$ flux in the $y$ direction by $2\pi\left(K^{-1}\right)^{jl}$ for all $l$. A diagonal layer $\Omega_\text{d}^{k,s}$ is coupled to $A^k$ only, and contributes a Berry phase of
\begin{equation*}
    \theta_\text{d,ij}^k=-2\pi\,\text{sgn}(\Delta_k)\left(K^{-1}\right)^{ik}\left(K^{-1}\right)^{jk},
\end{equation*}
whereas an off-diagonal layer $\Omega_\text{o}^{kl,t}$, $k<l$, is coupled to $A^k$ and $A^l$, and contributes
\begin{multline*}
    \theta_\text{o,ij}^{kl}=-2\pi\,\text{sgn}(K_{kl})\left[\left(K^{-1}\right)^{ik}+\left(K^{-1}\right)^{il}\right]\\
    \times\left[\left(K^{-1}\right)^{jk}+\left(K^{-1}\right)^{jl}\right].
\end{multline*}
The braiding phase of $\mathcal{W}_2^i\mathcal{W}_3^i$ with $\mathcal{W}_2^j\mathcal{W}_3^j$ is then
\begin{equation*}
    \sum_k |\Delta_k|\theta_\text{d,ij}^k+\sum_{k<l} |K_{kl}|\theta_\text{o,ij}^{kl}=-2\pi\left(K^{-1}\right)^{ij},
\end{equation*}
as can be confirmed by a straightforward calculation. Finally, we find the total braiding phase to be
\begin{equation*}
    2\pi \left(K^{-1}\right)^{ij}+2\pi \left(K^{-1}\right)^{ij}-2\pi \left(K^{-1}\right)^{ij}=2\pi \left(K^{-1}\right)^{ij},
\end{equation*}
as expected.

The string operators allow us to understand the profile of fractional excitations in the $z$ direction, which is determined by the fractional part of $K^{-1}$ (the integral part of $K^{-1}$ corresponds to local fermion and integer flux excitations). In particular, for the example of $K_{\text{nF}}$, the entries of each row of $K_{\text{nF}}^{-1}$ decay exponentially, which means that both $\mathcal{W}^i_2$ and $\mathcal{W}^i_3$ become exponentially close to the identity as we move in the $z$ direction ($\mathcal{W}^i_1$ is always local in the $z$ direction). Therefore, $\mathcal{W}^i$ is local in the $z$ direction with an exponentially decaying tail, and the fractional excitations are localized particles. On the other hand, for $K_{\text{gl}}$, the fractional parts of the entries of $K_{\text{gl}}^{-1}$ do not decay. This means that the fractional excitations in the $K_{\text{gl}}$ theory, if valid at all, are line excitations extended along the $z$ direction.

\section{Discussion}
\label{sec:conclusion}

In this paper, we established the iCS theory as a viable path to study a variety of fractonic phases. First of all, we showed in section~\ref{subsec:model} that iCS theories with a quasi-diagonal $K$ matrix are indeed legitimate local $3+1$D models by giving an explicit lattice realization for the theory. Using the method discussed in section~\ref{sec:spectrum}, we can further determine which iCS theories are gapped and which are gapless. Moreover, we found in section~\ref{subsec:string} the explicit form of the string operators that create the fractional excitations in the model. From the string operators, we can learn about the nature of the fractional excitations (for example when they are localized point excitations versus when they are extensive line excitations). Based on these understandings, we found through examples an interesting variety of fractonic phenomena in iCS theories with quasi-diagonal $K$ matrices. There are 1-foliated fracton models; there are abelian type~I models which do not have a foliation structure --- a feature not present in previously studied models; and there are gapless theories whose nature is not clear. Some of the non-foliated gapped models have been studied previously from the perspective of coupled fractional quantum Hall layers\cite{Qiu90,Qiu90-2,Naud00,Naud01}, which interestingly suggests a route toward experimental realization of these particular fracton phases.

The next step would be to study iCS theories more systematically and address questions such as
\begin{enumerate}[leftmargin=*]
    \item For gapped iCS theories, how can foliated theories be distinguished from non-foliated ones?
    \item If an iCS theory is foliated, how does one find the RG procedure that extracts 2D layers?
    \item How can we understand the non-foliated models, for example in terms of RG $s$-sourcery?\cite{SwingleSSource}
    \item What is the nature of the gapless models?
\end{enumerate}
We hope that by addressing these questions, we can have a more complete picture of possible fractonic orders, beyond what we can learn from exactly solvable models or other frameworks. 

Of course, the possibilities represented by the iCS theories are limited. The only kind of fractional excitations in these models are planons in the $xy$ plane. They do not even contain fracton excitations which are completely immobile. But as we have learned from previous studies, planons play an important role in type~I models. Once we have a better understanding of planons, maybe we can combine them with fractons and other sub-dimensional excitations to achieve a more complete story. 

\begin{acknowledgments}
We are indebted to inspiring discussions with Tina Zhang, Xiao-Gang Wen, Yuan-Ming Lu, Zhenghan Wang, and Kevin Slagle. We also thank Po-Shen Hsin for pointing out that a fermionic abelian topological order can always be decomposed into a bosonic abelian topological order and transparent fermions. X. M, W.S. and X.C. are supported by the National Science Foundation under award number DMR-1654340, the Simons collaboration on ``Ultra-Quantum Matter'' and the Institute for Quantum Information and Matter at Caltech. X.C. is also supported by the Walter Burke Institute for Theoretical Physics at Caltech. M. C. is supported by NSF CAREER (DMR-1846109) and the Alfred P. Sloan foundation. This work was supported in part by
funds provided by the U.S. Department of Energy
(D.O.E.) under cooperative research agreement 
DE-SC0009919, by the Simons Collaboration on Ultra-Quantum Matter, which is a grant from the Simons Foundation (651440, XC, ML, JM, XM, WS).
\end{acknowledgments}

\bibliography{references}

\begin{thebibliography}{38}%
\makeatletter
\providecommand \@ifxundefined [1]{%
 \@ifx{#1\undefined}
}%
\providecommand \@ifnum [1]{%
 \ifnum #1\expandafter \@firstoftwo
 \else \expandafter \@secondoftwo
 \fi
}%
\providecommand \@ifx [1]{%
 \ifx #1\expandafter \@firstoftwo
 \else \expandafter \@secondoftwo
 \fi
}%
\providecommand \natexlab [1]{#1}%
\providecommand \enquote  [1]{``#1''}%
\providecommand \bibnamefont  [1]{#1}%
\providecommand \bibfnamefont [1]{#1}%
\providecommand \citenamefont [1]{#1}%
\providecommand \href@noop [0]{\@secondoftwo}%
\providecommand \href [0]{\begingroup \@sanitize@url \@href}%
\providecommand \@href[1]{\@@startlink{#1}\@@href}%
\providecommand \@@href[1]{\endgroup#1\@@endlink}%
\providecommand \@sanitize@url [0]{\catcode `\\12\catcode `\$12\catcode
  `\&12\catcode `\#12\catcode `\^12\catcode `\_12\catcode `\%12\relax}%
\providecommand \@@startlink[1]{}%
\providecommand \@@endlink[0]{}%
\providecommand \url  [0]{\begingroup\@sanitize@url \@url }%
\providecommand \@url [1]{\endgroup\@href {#1}{\urlprefix }}%
\providecommand \urlprefix  [0]{URL }%
\providecommand \Eprint [0]{\href }%
\providecommand \doibase [0]{http://dx.doi.org/}%
\providecommand \selectlanguage [0]{\@gobble}%
\providecommand \bibinfo  [0]{\@secondoftwo}%
\providecommand \bibfield  [0]{\@secondoftwo}%
\providecommand \translation [1]{[#1]}%
\providecommand \BibitemOpen [0]{}%
\providecommand \bibitemStop [0]{}%
\providecommand \bibitemNoStop [0]{.\EOS\space}%
\providecommand \EOS [0]{\spacefactor3000\relax}%
\providecommand \BibitemShut  [1]{\csname bibitem#1\endcsname}%
\let\auto@bib@innerbib\@empty
\bibitem [{\citenamefont {Nandkishore}\ and\ \citenamefont
  {Hermele}(2019)}]{Nandkishore_2019}%
  \BibitemOpen
  \bibfield  {author} {\bibinfo {author} {\bibfnamefont {R.~M.}\ \bibnamefont
  {Nandkishore}}\ and\ \bibinfo {author} {\bibfnamefont {M.}~\bibnamefont
  {Hermele}},\ }\href {\doibase 10.1146/annurev-conmatphys-031218-013604}
  {\bibfield  {journal} {\bibinfo  {journal} {Annual Review of Condensed Matter
  Physics}\ }\textbf {\bibinfo {volume} {10}},\ \bibinfo {pages} {295}
  (\bibinfo {year} {2019})}\BibitemShut {NoStop}%
\bibitem [{\citenamefont {Pretko}\ \emph {et~al.}(2020)\citenamefont {Pretko},
  \citenamefont {Chen},\ and\ \citenamefont {You}}]{Pretko_2020}%
  \BibitemOpen
  \bibfield  {author} {\bibinfo {author} {\bibfnamefont {M.}~\bibnamefont
  {Pretko}}, \bibinfo {author} {\bibfnamefont {X.}~\bibnamefont {Chen}}, \ and\
  \bibinfo {author} {\bibfnamefont {Y.}~\bibnamefont {You}},\ }\href {\doibase
  10.1142/s0217751x20300033} {\bibfield  {journal} {\bibinfo  {journal}
  {International Journal of Modern Physics A}\ }\textbf {\bibinfo {volume}
  {35}},\ \bibinfo {pages} {2030003} (\bibinfo {year} {2020})}\BibitemShut
  {NoStop}%
\bibitem [{\citenamefont {Chamon}(2005)}]{ChamonModel}%
  \BibitemOpen
  \bibfield  {author} {\bibinfo {author} {\bibfnamefont {C.}~\bibnamefont
  {Chamon}},\ }\href {\doibase 10.1103/PhysRevLett.94.040402} {\bibfield
  {journal} {\bibinfo  {journal} {Phys. Rev. Lett.}\ }\textbf {\bibinfo
  {volume} {94}},\ \bibinfo {pages} {040402} (\bibinfo {year}
  {2005})}\BibitemShut {NoStop}%
\bibitem [{\citenamefont {Haah}(2011)}]{HaahCode}%
  \BibitemOpen
  \bibfield  {author} {\bibinfo {author} {\bibfnamefont {J.}~\bibnamefont
  {Haah}},\ }\href {\doibase 10.1103/PhysRevA.83.042330} {\bibfield  {journal}
  {\bibinfo  {journal} {Phys. Rev. A}\ }\textbf {\bibinfo {volume} {83}},\
  \bibinfo {pages} {042330} (\bibinfo {year} {2011})}\BibitemShut {NoStop}%
\bibitem [{\citenamefont {Vijay}\ \emph {et~al.}(2016)\citenamefont {Vijay},
  \citenamefont {Haah},\ and\ \citenamefont {Fu}}]{Sagar16}%
  \BibitemOpen
  \bibfield  {author} {\bibinfo {author} {\bibfnamefont {S.}~\bibnamefont
  {Vijay}}, \bibinfo {author} {\bibfnamefont {J.}~\bibnamefont {Haah}}, \ and\
  \bibinfo {author} {\bibfnamefont {L.}~\bibnamefont {Fu}},\ }\href {\doibase
  10.1103/PhysRevB.94.235157} {\bibfield  {journal} {\bibinfo  {journal} {Phys.
  Rev. B}\ }\textbf {\bibinfo {volume} {94}},\ \bibinfo {pages} {235157}
  (\bibinfo {year} {2016})}\BibitemShut {NoStop}%
\bibitem [{\citenamefont {Vijay}\ \emph {et~al.}(2015)\citenamefont {Vijay},
  \citenamefont {Haah},\ and\ \citenamefont {Fu}}]{VijayFracton}%
  \BibitemOpen
  \bibfield  {author} {\bibinfo {author} {\bibfnamefont {S.}~\bibnamefont
  {Vijay}}, \bibinfo {author} {\bibfnamefont {J.}~\bibnamefont {Haah}}, \ and\
  \bibinfo {author} {\bibfnamefont {L.}~\bibnamefont {Fu}},\ }\href {\doibase
  10.1103/PhysRevB.92.235136} {\bibfield  {journal} {\bibinfo  {journal} {Phys.
  Rev. B}\ }\textbf {\bibinfo {volume} {92}},\ \bibinfo {pages} {235136}
  (\bibinfo {year} {2015})}\BibitemShut {NoStop}%
\bibitem [{\citenamefont {Prem}\ \emph {et~al.}(2019)\citenamefont {Prem},
  \citenamefont {Huang}, \citenamefont {Song},\ and\ \citenamefont
  {Hermele}}]{Prem_2019}%
  \BibitemOpen
  \bibfield  {author} {\bibinfo {author} {\bibfnamefont {A.}~\bibnamefont
  {Prem}}, \bibinfo {author} {\bibfnamefont {S.-J.}\ \bibnamefont {Huang}},
  \bibinfo {author} {\bibfnamefont {H.}~\bibnamefont {Song}}, \ and\ \bibinfo
  {author} {\bibfnamefont {M.}~\bibnamefont {Hermele}},\ }\href {\doibase
  10.1103/physrevx.9.021010} {\bibfield  {journal} {\bibinfo  {journal}
  {Physical Review X}\ }\textbf {\bibinfo {volume} {9}} (\bibinfo {year}
  {2019}),\ 10.1103/physrevx.9.021010}\BibitemShut {NoStop}%
\bibitem [{\citenamefont {Song}\ \emph {et~al.}(2019)\citenamefont {Song},
  \citenamefont {Prem}, \citenamefont {Huang},\ and\ \citenamefont
  {Martin-Delgado}}]{Song2019}%
  \BibitemOpen
  \bibfield  {author} {\bibinfo {author} {\bibfnamefont {H.}~\bibnamefont
  {Song}}, \bibinfo {author} {\bibfnamefont {A.}~\bibnamefont {Prem}}, \bibinfo
  {author} {\bibfnamefont {S.-J.}\ \bibnamefont {Huang}}, \ and\ \bibinfo
  {author} {\bibfnamefont {M.~A.}\ \bibnamefont {Martin-Delgado}},\ }\href
  {\doibase 10.1103/PhysRevB.99.155118} {\bibfield  {journal} {\bibinfo
  {journal} {Phys. Rev. B}\ }\textbf {\bibinfo {volume} {99}},\ \bibinfo
  {pages} {155118} (\bibinfo {year} {2019})}\BibitemShut {NoStop}%
\bibitem [{\citenamefont {Ma}\ \emph {et~al.}(2018)\citenamefont {Ma},
  \citenamefont {Schmitz}, \citenamefont {Parameswaran}, \citenamefont
  {Hermele},\ and\ \citenamefont {Nandkishore}}]{HermeleEntropy}%
  \BibitemOpen
  \bibfield  {author} {\bibinfo {author} {\bibfnamefont {H.}~\bibnamefont
  {Ma}}, \bibinfo {author} {\bibfnamefont {A.~T.}\ \bibnamefont {Schmitz}},
  \bibinfo {author} {\bibfnamefont {S.~A.}\ \bibnamefont {Parameswaran}},
  \bibinfo {author} {\bibfnamefont {M.}~\bibnamefont {Hermele}}, \ and\
  \bibinfo {author} {\bibfnamefont {R.~M.}\ \bibnamefont {Nandkishore}},\
  }\href {\doibase 10.1103/PhysRevB.97.125101} {\bibfield  {journal} {\bibinfo
  {journal} {Phys. Rev. B}\ }\textbf {\bibinfo {volume} {97}},\ \bibinfo
  {pages} {125101} (\bibinfo {year} {2018})}\BibitemShut {NoStop}%
\bibitem [{\citenamefont {Shi}\ and\ \citenamefont {Lu}(2018)}]{ShiEntropy}%
  \BibitemOpen
  \bibfield  {author} {\bibinfo {author} {\bibfnamefont {B.}~\bibnamefont
  {Shi}}\ and\ \bibinfo {author} {\bibfnamefont {Y.-M.}\ \bibnamefont {Lu}},\
  }\href {\doibase 10.1103/PhysRevB.97.144106} {\bibfield  {journal} {\bibinfo
  {journal} {Phys. Rev. B}\ }\textbf {\bibinfo {volume} {97}},\ \bibinfo
  {pages} {144106} (\bibinfo {year} {2018})}\BibitemShut {NoStop}%
\bibitem [{\citenamefont {He}\ \emph {et~al.}(2018)\citenamefont {He},
  \citenamefont {Zheng}, \citenamefont {Bernevig},\ and\ \citenamefont
  {Regnault}}]{BernevigEntropy}%
  \BibitemOpen
  \bibfield  {author} {\bibinfo {author} {\bibfnamefont {H.}~\bibnamefont
  {He}}, \bibinfo {author} {\bibfnamefont {Y.}~\bibnamefont {Zheng}}, \bibinfo
  {author} {\bibfnamefont {B.~A.}\ \bibnamefont {Bernevig}}, \ and\ \bibinfo
  {author} {\bibfnamefont {N.}~\bibnamefont {Regnault}},\ }\href {\doibase
  10.1103/PhysRevB.97.125102} {\bibfield  {journal} {\bibinfo  {journal} {Phys.
  Rev. B}\ }\textbf {\bibinfo {volume} {97}},\ \bibinfo {pages} {125102}
  (\bibinfo {year} {2018})}\BibitemShut {NoStop}%
\bibitem [{\citenamefont {Shirley}\ \emph
  {et~al.}(2019{\natexlab{a}})\citenamefont {Shirley}, \citenamefont {Slagle},\
  and\ \citenamefont {Chen}}]{Shirley_2019}%
  \BibitemOpen
  \bibfield  {author} {\bibinfo {author} {\bibfnamefont {W.}~\bibnamefont
  {Shirley}}, \bibinfo {author} {\bibfnamefont {K.}~\bibnamefont {Slagle}}, \
  and\ \bibinfo {author} {\bibfnamefont {X.}~\bibnamefont {Chen}},\ }\href
  {\doibase 10.21468/scipostphys.6.1.015} {\bibfield  {journal} {\bibinfo
  {journal} {SciPost Physics}\ }\textbf {\bibinfo {volume} {6}} (\bibinfo
  {year} {2019}{\natexlab{a}}),\ 10.21468/scipostphys.6.1.015}\BibitemShut
  {NoStop}%
\bibitem [{\citenamefont {Shirley}\ \emph {et~al.}(2018)\citenamefont
  {Shirley}, \citenamefont {Slagle}, \citenamefont {Wang},\ and\ \citenamefont
  {Chen}}]{Shirley_2018}%
  \BibitemOpen
  \bibfield  {author} {\bibinfo {author} {\bibfnamefont {W.}~\bibnamefont
  {Shirley}}, \bibinfo {author} {\bibfnamefont {K.}~\bibnamefont {Slagle}},
  \bibinfo {author} {\bibfnamefont {Z.}~\bibnamefont {Wang}}, \ and\ \bibinfo
  {author} {\bibfnamefont {X.}~\bibnamefont {Chen}},\ }\href {\doibase
  10.1103/physrevx.8.031051} {\bibfield  {journal} {\bibinfo  {journal}
  {Physical Review X}\ }\textbf {\bibinfo {volume} {8}} (\bibinfo {year}
  {2018}),\ 10.1103/physrevx.8.031051}\BibitemShut {NoStop}%
\bibitem [{\citenamefont {Pretko}(2017{\natexlab{a}})}]{PretkoU1}%
  \BibitemOpen
  \bibfield  {author} {\bibinfo {author} {\bibfnamefont {M.}~\bibnamefont
  {Pretko}},\ }\href {\doibase 10.1103/PhysRevB.95.115139} {\bibfield
  {journal} {\bibinfo  {journal} {Phys. Rev. B}\ }\textbf {\bibinfo {volume}
  {95}},\ \bibinfo {pages} {115139} (\bibinfo {year}
  {2017}{\natexlab{a}})}\BibitemShut {NoStop}%
\bibitem [{\citenamefont
  {Pretko}(2017{\natexlab{b}})}]{electromagnetismPretko}%
  \BibitemOpen
  \bibfield  {author} {\bibinfo {author} {\bibfnamefont {M.}~\bibnamefont
  {Pretko}},\ }\href {\doibase 10.1103/PhysRevB.96.035119} {\bibfield
  {journal} {\bibinfo  {journal} {Phys. Rev. B}\ }\textbf {\bibinfo {volume}
  {96}},\ \bibinfo {pages} {035119} (\bibinfo {year}
  {2017}{\natexlab{b}})}\BibitemShut {NoStop}%
\bibitem [{\citenamefont {Gromov}(2019)}]{Gromov2019}%
  \BibitemOpen
  \bibfield  {author} {\bibinfo {author} {\bibfnamefont {A.}~\bibnamefont
  {Gromov}},\ }\href {\doibase 10.1103/PhysRevX.9.031035} {\bibfield  {journal}
  {\bibinfo  {journal} {Phys. Rev. X}\ }\textbf {\bibinfo {volume} {9}},\
  \bibinfo {pages} {031035} (\bibinfo {year} {2019})}\BibitemShut {NoStop}%
\bibitem [{\citenamefont {Seiberg}\ and\ \citenamefont
  {Shao}(2020)}]{seiberg2020exotic}%
  \BibitemOpen
  \bibfield  {author} {\bibinfo {author} {\bibfnamefont {N.}~\bibnamefont
  {Seiberg}}\ and\ \bibinfo {author} {\bibfnamefont {S.-H.}\ \bibnamefont
  {Shao}},\ }\href@noop {} {\enquote {\bibinfo {title} {Exotic $u(1)$
  symmetries, duality, and fractons in 3+1-dimensional quantum field theory},}\
  } (\bibinfo {year} {2020}),\ \Eprint {http://arxiv.org/abs/2004.00015}
  {arXiv:2004.00015 [cond-mat.str-el]} \BibitemShut {NoStop}%
\bibitem [{\citenamefont {Wen}(1990)}]{WenRigid}%
  \BibitemOpen
  \bibfield  {author} {\bibinfo {author} {\bibfnamefont {X.~G.}\ \bibnamefont
  {Wen}},\ }\href {\doibase 10.1142/S0217979290000139} {\bibfield  {journal}
  {\bibinfo  {journal} {International Journal of Modern Physics B}\ }\textbf
  {\bibinfo {volume} {04}},\ \bibinfo {pages} {239} (\bibinfo {year}
  {1990})}\BibitemShut {NoStop}%
\bibitem [{\citenamefont {Wen}\ and\ \citenamefont
  {Zee}(1992)}]{WenChernSimons}%
  \BibitemOpen
  \bibfield  {author} {\bibinfo {author} {\bibfnamefont {X.~G.}\ \bibnamefont
  {Wen}}\ and\ \bibinfo {author} {\bibfnamefont {A.}~\bibnamefont {Zee}},\
  }\href {\doibase 10.1103/PhysRevB.46.2290} {\bibfield  {journal} {\bibinfo
  {journal} {Phys. Rev. B}\ }\textbf {\bibinfo {volume} {46}},\ \bibinfo
  {pages} {2290} (\bibinfo {year} {1992})}\BibitemShut {NoStop}%
\bibitem [{\citenamefont {Shirley}\ \emph
  {et~al.}(2019{\natexlab{b}})\citenamefont {Shirley}, \citenamefont {Slagle},\
  and\ \citenamefont {Chen}}]{Shirley2019twisted}%
  \BibitemOpen
  \bibfield  {author} {\bibinfo {author} {\bibfnamefont {W.}~\bibnamefont
  {Shirley}}, \bibinfo {author} {\bibfnamefont {K.}~\bibnamefont {Slagle}}, \
  and\ \bibinfo {author} {\bibfnamefont {X.}~\bibnamefont {Chen}},\ }\href@noop
  {} {\enquote {\bibinfo {title} {Twisted foliated fracton phases},}\ }
  (\bibinfo {year} {2019}{\natexlab{b}}),\ \Eprint
  {http://arxiv.org/abs/1907.09048} {arXiv:1907.09048 [cond-mat.str-el]}
  \BibitemShut {NoStop}%
\bibitem [{\citenamefont {Qiu}\ \emph {et~al.}(1989)\citenamefont {Qiu},
  \citenamefont {Joynt},\ and\ \citenamefont {MacDonald}}]{qiu1989phases}%
  \BibitemOpen
  \bibfield  {author} {\bibinfo {author} {\bibfnamefont {X.}~\bibnamefont
  {Qiu}}, \bibinfo {author} {\bibfnamefont {R.}~\bibnamefont {Joynt}}, \ and\
  \bibinfo {author} {\bibfnamefont {A.}~\bibnamefont {MacDonald}},\ }\href
  {https://doi.org/10.1103/PhysRevB.40.11943} {\bibfield  {journal} {\bibinfo
  {journal} {Physical Review B}\ }\textbf {\bibinfo {volume} {40}},\ \bibinfo
  {pages} {11943} (\bibinfo {year} {1989})}\BibitemShut {NoStop}%
\bibitem [{\citenamefont {Qiu}\ \emph {et~al.}(1990{\natexlab{a}})\citenamefont
  {Qiu}, \citenamefont {Joynt},\ and\ \citenamefont {MacDonald}}]{Qiu90-2}%
  \BibitemOpen
  \bibfield  {author} {\bibinfo {author} {\bibfnamefont {X.}~\bibnamefont
  {Qiu}}, \bibinfo {author} {\bibfnamefont {R.}~\bibnamefont {Joynt}}, \ and\
  \bibinfo {author} {\bibfnamefont {A.~H.}\ \bibnamefont {MacDonald}},\ }\href
  {\doibase 10.1103/PhysRevB.42.1339} {\bibfield  {journal} {\bibinfo
  {journal} {Phys. Rev. B}\ }\textbf {\bibinfo {volume} {42}},\ \bibinfo
  {pages} {1339} (\bibinfo {year} {1990}{\natexlab{a}})}\BibitemShut {NoStop}%
\bibitem [{\citenamefont {Naud}\ \emph {et~al.}(2000)\citenamefont {Naud},
  \citenamefont {Pryadko},\ and\ \citenamefont {Sondhi}}]{Naud00}%
  \BibitemOpen
  \bibfield  {author} {\bibinfo {author} {\bibfnamefont {J.~D.}\ \bibnamefont
  {Naud}}, \bibinfo {author} {\bibfnamefont {L.~P.}\ \bibnamefont {Pryadko}}, \
  and\ \bibinfo {author} {\bibfnamefont {S.~L.}\ \bibnamefont {Sondhi}},\
  }\href {\doibase 10.1103/PhysRevLett.85.5408} {\bibfield  {journal} {\bibinfo
   {journal} {Phys. Rev. Lett.}\ }\textbf {\bibinfo {volume} {85}},\ \bibinfo
  {pages} {5408} (\bibinfo {year} {2000})}\BibitemShut {NoStop}%
\bibitem [{\citenamefont {Naud}\ \emph {et~al.}(2001)\citenamefont {Naud},
  \citenamefont {Pryadko},\ and\ \citenamefont {Sondhi}}]{Naud01}%
  \BibitemOpen
  \bibfield  {author} {\bibinfo {author} {\bibfnamefont {J.}~\bibnamefont
  {Naud}}, \bibinfo {author} {\bibfnamefont {L.~P.}\ \bibnamefont {Pryadko}}, \
  and\ \bibinfo {author} {\bibfnamefont {S.}~\bibnamefont {Sondhi}},\ }\href
  {\doibase https://doi.org/10.1016/S0550-3213(00)00679-9} {\bibfield
  {journal} {\bibinfo  {journal} {Nuclear Physics B}\ }\textbf {\bibinfo
  {volume} {594}},\ \bibinfo {pages} {713 } (\bibinfo {year}
  {2001})}\BibitemShut {NoStop}%
\bibitem [{\citenamefont {Wall}(1963)}]{WALL1963281}%
  \BibitemOpen
  \bibfield  {author} {\bibinfo {author} {\bibfnamefont {C.}~\bibnamefont
  {Wall}},\ }\href {\doibase https://doi.org/10.1016/0040-9383(63)90012-0}
  {\bibfield  {journal} {\bibinfo  {journal} {Topology}\ }\textbf {\bibinfo
  {volume} {2}},\ \bibinfo {pages} {281 } (\bibinfo {year} {1963})}\BibitemShut
  {NoStop}%
\bibitem [{\citenamefont {Wang}\ and\ \citenamefont
  {Wang}(2020)}]{wang2020abelian}%
  \BibitemOpen
  \bibfield  {author} {\bibinfo {author} {\bibfnamefont {L.}~\bibnamefont
  {Wang}}\ and\ \bibinfo {author} {\bibfnamefont {Z.}~\bibnamefont {Wang}},\
  }\href@noop {} {\enquote {\bibinfo {title} {In and around abelian anyon
  models},}\ } (\bibinfo {year} {2020}),\ \Eprint
  {http://arxiv.org/abs/2004.12048} {arXiv:2004.12048 [math.QA]} \BibitemShut
  {NoStop}%
\bibitem [{\citenamefont {Levin}\ and\ \citenamefont
  {Fisher}(2009)}]{Levin2009}%
  \BibitemOpen
  \bibfield  {author} {\bibinfo {author} {\bibfnamefont {M.}~\bibnamefont
  {Levin}}\ and\ \bibinfo {author} {\bibfnamefont {M.~P.~A.}\ \bibnamefont
  {Fisher}},\ }\href {\doibase 10.1103/PhysRevB.79.235315} {\bibfield
  {journal} {\bibinfo  {journal} {Phys. Rev. B}\ }\textbf {\bibinfo {volume}
  {79}},\ \bibinfo {pages} {235315} (\bibinfo {year} {2009})}\BibitemShut
  {NoStop}%
\bibitem [{\citenamefont {Ma}\ \emph {et~al.}(2017)\citenamefont {Ma},
  \citenamefont {Lake}, \citenamefont {Chen},\ and\ \citenamefont
  {Hermele}}]{MaLayers}%
  \BibitemOpen
  \bibfield  {author} {\bibinfo {author} {\bibfnamefont {H.}~\bibnamefont
  {Ma}}, \bibinfo {author} {\bibfnamefont {E.}~\bibnamefont {Lake}}, \bibinfo
  {author} {\bibfnamefont {X.}~\bibnamefont {Chen}}, \ and\ \bibinfo {author}
  {\bibfnamefont {M.}~\bibnamefont {Hermele}},\ }\href {\doibase
  10.1103/PhysRevB.95.245126} {\bibfield  {journal} {\bibinfo  {journal} {Phys.
  Rev. B}\ }\textbf {\bibinfo {volume} {95}},\ \bibinfo {pages} {245126}
  (\bibinfo {year} {2017})}\BibitemShut {NoStop}%
\bibitem [{\citenamefont {Shirley}\ \emph
  {et~al.}(2019{\natexlab{c}})\citenamefont {Shirley}, \citenamefont {Slagle},\
  and\ \citenamefont {Chen}}]{Shirley_2019_excitation}%
  \BibitemOpen
  \bibfield  {author} {\bibinfo {author} {\bibfnamefont {W.}~\bibnamefont
  {Shirley}}, \bibinfo {author} {\bibfnamefont {K.}~\bibnamefont {Slagle}}, \
  and\ \bibinfo {author} {\bibfnamefont {X.}~\bibnamefont {Chen}},\ }\href
  {\doibase 10.1016/j.aop.2019.167922} {\bibfield  {journal} {\bibinfo
  {journal} {Annals of Physics}\ }\textbf {\bibinfo {volume} {410}},\ \bibinfo
  {pages} {167922} (\bibinfo {year} {2019}{\natexlab{c}})}\BibitemShut
  {NoStop}%
\bibitem [{\citenamefont {Shirley}\ \emph
  {et~al.}(2019{\natexlab{d}})\citenamefont {Shirley}, \citenamefont {Slagle},\
  and\ \citenamefont {Chen}}]{Shirley_2019_checkerboard}%
  \BibitemOpen
  \bibfield  {author} {\bibinfo {author} {\bibfnamefont {W.}~\bibnamefont
  {Shirley}}, \bibinfo {author} {\bibfnamefont {K.}~\bibnamefont {Slagle}}, \
  and\ \bibinfo {author} {\bibfnamefont {X.}~\bibnamefont {Chen}},\ }\href
  {\doibase 10.1103/physrevb.99.115123} {\bibfield  {journal} {\bibinfo
  {journal} {Physical Review B}\ }\textbf {\bibinfo {volume} {99}} (\bibinfo
  {year} {2019}{\natexlab{d}}),\ 10.1103/physrevb.99.115123}\BibitemShut
  {NoStop}%
\bibitem [{\citenamefont {Wang}\ \emph {et~al.}(2019)\citenamefont {Wang},
  \citenamefont {Shirley},\ and\ \citenamefont {Chen}}]{Wang_2019}%
  \BibitemOpen
  \bibfield  {author} {\bibinfo {author} {\bibfnamefont {T.}~\bibnamefont
  {Wang}}, \bibinfo {author} {\bibfnamefont {W.}~\bibnamefont {Shirley}}, \
  and\ \bibinfo {author} {\bibfnamefont {X.}~\bibnamefont {Chen}},\ }\href
  {\doibase 10.1103/physrevb.100.085127} {\bibfield  {journal} {\bibinfo
  {journal} {Physical Review B}\ }\textbf {\bibinfo {volume} {100}} (\bibinfo
  {year} {2019}),\ 10.1103/physrevb.100.085127}\BibitemShut {NoStop}%
\bibitem [{\citenamefont {Vidal}(2007)}]{VidalRG}%
  \BibitemOpen
  \bibfield  {author} {\bibinfo {author} {\bibfnamefont {G.}~\bibnamefont
  {Vidal}},\ }\href {\doibase 10.1103/PhysRevLett.99.220405} {\bibfield
  {journal} {\bibinfo  {journal} {Phys. Rev. Lett.}\ }\textbf {\bibinfo
  {volume} {99}},\ \bibinfo {pages} {220405} (\bibinfo {year}
  {2007})}\BibitemShut {NoStop}%
\bibitem [{\citenamefont {Qiu}\ \emph {et~al.}(1990{\natexlab{b}})\citenamefont
  {Qiu}, \citenamefont {Joynt},\ and\ \citenamefont {MacDonald}}]{Qiu90}%
  \BibitemOpen
  \bibfield  {author} {\bibinfo {author} {\bibfnamefont {X.}~\bibnamefont
  {Qiu}}, \bibinfo {author} {\bibfnamefont {R.}~\bibnamefont {Joynt}}, \ and\
  \bibinfo {author} {\bibfnamefont {A.~H.}\ \bibnamefont {MacDonald}},\ }\href
  {\doibase 10.1103/PhysRevB.42.1339} {\bibfield  {journal} {\bibinfo
  {journal} {Phys. Rev. B}\ }\textbf {\bibinfo {volume} {42}},\ \bibinfo
  {pages} {1339} (\bibinfo {year} {1990}{\natexlab{b}})}\BibitemShut {NoStop}%
\bibitem [{Note1()}]{Note1}%
  \BibitemOpen
  \bibinfo {note} {The integers 1, 4, 11, 29, etc. form a sequence known as the
  Lucas numbers.}\BibitemShut {Stop}%
\bibitem [{\citenamefont {Kitaev}(2006)}]{KitaevAnyons}%
  \BibitemOpen
  \bibfield  {author} {\bibinfo {author} {\bibfnamefont {A.}~\bibnamefont
  {Kitaev}},\ }\href {\doibase 10.1016/j.aop.2005.10.005} {\bibfield  {journal}
  {\bibinfo  {journal} {Annals of Physics}\ }\textbf {\bibinfo {volume}
  {321}},\ \bibinfo {pages} {2 } (\bibinfo {year} {2006})},\ \bibinfo {note}
  {january Special Issue}\BibitemShut {NoStop}%
\bibitem [{\citenamefont {Kalmeyer}\ and\ \citenamefont
  {Laughlin}(1987)}]{Kalmeyer87}%
  \BibitemOpen
  \bibfield  {author} {\bibinfo {author} {\bibfnamefont {V.}~\bibnamefont
  {Kalmeyer}}\ and\ \bibinfo {author} {\bibfnamefont {R.~B.}\ \bibnamefont
  {Laughlin}},\ }\href {\doibase 10.1103/PhysRevLett.59.2095} {\bibfield
  {journal} {\bibinfo  {journal} {Phys. Rev. Lett.}\ }\textbf {\bibinfo
  {volume} {59}},\ \bibinfo {pages} {2095} (\bibinfo {year}
  {1987})}\BibitemShut {NoStop}%
\bibitem [{\citenamefont {Hastings}\ and\ \citenamefont
  {Wen}(2005)}]{Hastings2005}%
  \BibitemOpen
  \bibfield  {author} {\bibinfo {author} {\bibfnamefont {M.~B.}\ \bibnamefont
  {Hastings}}\ and\ \bibinfo {author} {\bibfnamefont {X.-G.}\ \bibnamefont
  {Wen}},\ }\href {\doibase 10.1103/PhysRevB.72.045141} {\bibfield  {journal}
  {\bibinfo  {journal} {Phys. Rev. B}\ }\textbf {\bibinfo {volume} {72}},\
  \bibinfo {pages} {045141} (\bibinfo {year} {2005})}\BibitemShut {NoStop}%
\bibitem [{\citenamefont {Swingle}\ and\ \citenamefont
  {McGreevy}(2016)}]{SwingleSSource}%
  \BibitemOpen
  \bibfield  {author} {\bibinfo {author} {\bibfnamefont {B.}~\bibnamefont
  {Swingle}}\ and\ \bibinfo {author} {\bibfnamefont {J.}~\bibnamefont
  {McGreevy}},\ }\href {\doibase 10.1103/PhysRevB.93.045127} {\bibfield
  {journal} {\bibinfo  {journal} {Phys. Rev. B}\ }\textbf {\bibinfo {volume}
  {93}},\ \bibinfo {pages} {045127} (\bibinfo {year} {2016})}\BibitemShut
  {NoStop}%
\end{thebibliography}%

\appendix

\section{A class of foliated iCS theories}
\label{sec:KFnr}

The twisted 1-foliated fractonic order Eq.~\ref{eq:K_F} can be generalized to a class of foliated iCS theories, with $K$ matrix parameterized by integers $n$ and $r$:
\begin{equation*}
K_\text{F}(N)=
\begin{pmatrix}
\ddots\\
& 0 & n & -r\\
& n & 0\\
& -r & & 0 & n & -r\\
& & & n & 0\\
& & & -r & 0 & n\\
& & & & & n & 0\\
& & & & & & & \ddots
\end{pmatrix}_{4N\times 4N},
\end{equation*}
where $n>0$ and $r$ takes value in $\{0,1,...,n-1\}$. This $K_\text{F}$ can be obtained by boson condensation in a stack of $\mathbb{Z}_n\times\mathbb{Z}_n$ twisted gauge theories, in a way that naturally generalizes the procedure in Ref.~\onlinecite{Shirley2019twisted}. To see the foliation structure, take \[W=
\setcounter{MaxMatrixCols}{12}
\begin{pmatrix}
1 & & & & -1 & & & & 1\\
& 1\\
& & 1 & & & & -1\\
& & & 1 & & & & & & & & -1\\
& & & & 1\\
& & & & & 1 & & & & 1\\
& & & & & & 1\\
& & & 1 & & & & 1\\
& & & & 1 & & & & -1\\
& 1 & & & & & & & & -1\\
& & 1 & & & & -1 & & & & 1\\
& & & & & & & & & & & 1
\end{pmatrix},\]
where the action of $W$ outside this $12\times12$ block is the identity. We find that $WK_\text{F}(N)W^T$ decouples into $K_\text{F}(N-2)$ and two copies of $\mathbb{Z}_n\times\mathbb{Z}_n$ twisted gauge theories described by \[K_0=
\begin{pmatrix}
0 & n & -r\\
n & 0\\
-r & & 0 & n\\
& & n & 0
\end{pmatrix}.\]
The fusion group when $n$ and $r$ are coprime is
\begin{equation*}
    G=
    \begin{cases}
    \mathbb{Z}_{n^2}^{2N-2}\times\mathbb{Z}_n^4 & \mbox{if } N \mbox{ is even},\\
    \mathbb{Z}_{n^2}^{2N} & \mbox{if } N \mbox{ is odd, } n \mbox{ is odd,}\\
    \mathbb{Z}_{n^2}^{2N-2}\times\mathbb{Z}_{n^2/2}^2\times\mathbb{Z}_2^2 & \mbox{if } N \mbox{ if odd, } n \mbox{ is even.}
\end{cases}
\end{equation*}
When $n$ and $r$ are not coprime, we can factor out $\text{gcd}(n,r)$ from $K$ and analyze similarly.

\section{A class of non-foliated iCS theories}
\label{sec:det}

In this appendix, we generalize $K_{\text{nF}}$ and $K_{\text{gl}}$ to a class of non-foliated iCS theories described by
\begin{equation*}
    K(N)=
    \begin{pmatrix}
    n & 1 & & & 1\\
    1& n & 1 & & \\
    & \ddots & \ddots & \ddots & \\
    & & 1 & n & 1 \\
    1 & & & 1 & n  
    \end{pmatrix}_{N\times N},
\end{equation*}
$n\in\mathbb{Z}$, and derive various quantities regarding $K(N)$. We first consider a different matrix
\begin{equation*}
    K'(N)=
    \begin{pmatrix}
    n & 1 & & & \\ 1& n & 1 & & \\ & \ddots & \ddots & \ddots & \\ & & 1 & n & 1 \\ & & & 1 & n  
    \end{pmatrix}_{N\times N},
\end{equation*}
which is obtained from $K(N)$ by removing the entries in the top-right and bottom-left corners. We will compute $D'(N):=\det K'(N)$, which will be useful when we compute $D(N):=\det K(N)$ and $K(N)^{-1}$ later in this appendix. To do this, we need to take exactly one entry from each row and one from each column. If entry $(1,1)$ is used from the first row, then the rest is just $K'(N-1)$. Otherwise, entry $(1,2)$ must be used, and therefore so must be entry $(2,1)$, and the rest is $K'(N-2)$. We thus obtain the recurrence relation
\begin{equation}
\label{eq:D'_recur}
    D'(N)=nD'(N-1)-D'(N-2).
\end{equation}
Solving this with the initial conditions $D'(0)=1$ and $D'(1)=n$, we find
\begin{equation*}
    D'(N)=\frac{1}{\sqrt{n^2-4}}\left(x_+^{N+1}-x_-^{N+1}\right),
\end{equation*}
where $x_\pm=\left(n\pm\sqrt{n^2-4}\right)/2$.

Now we compute $D(N)$. Depending on whether entries $(1,N)$ and $(N,1)$ are used, we can write
\begin{equation}
\label{eq:D'_to_D}
    D(N)=D'(N)+(-1)^{N+1}+(-1)^{N+1}+D'(N-2),
\end{equation}
where the first term uses neither of the entries $(1,N)$ and $(N,1)$, the second and third terms uses exacly one, and the third term uses both. Further simplification then gives Eq.~\ref{eq:det(K)}. Incidentally, the recurrence relation Eq.~\ref{eq:D'_recur} has characteristic polynomial \[p'(x)=x^2-nx+1,\] and Eq.~\ref{eq:D'_to_D} implies that $D(N)$ satisfies a third order recurrence relation whose characteristic polynomial $p(x)$ has a third root $-1$ in addition to those of $\tilde{p}(x)$. Thus
\begin{align*}
    p(x)&=(x+1)p'(x)\\
    &=x^3-(n-1)x^2-(n-1)x+1,
\end{align*}
and $D(N)$ satisfies a third order recurrence relation \[D(N)=(n-1)D(N-1)+(n-1)D(N-2)-D(N-3).\]

The fusion group for $K(N)$ is of the form
\begin{equation*}
    G=G_1\times G_2=\mathbb{Z}_{a^{-1/2}D^{1/2}}\times\mathbb{Z}_{a^{1/2}D^{1/2}},
\end{equation*}
where $a$ depends on $n$ and $N$ as follows:
\begin{itemize}[leftmargin=*]
    \item If $N$ is odd, then $a=n+2$. A choice of generators is $(n+1)e_1+e_2$ for $G_1$ and $e_1$ for $G_2$.
    
    \item If $N$ is even and $n$ is odd, then $a=n^2-4$. A choice of generators is $\left(n(n+1)/2-2\right)e_1+e_2$ for $G_1$ and $e_1$ for $G_2$.
    
    \item If $N$ is even and $n$ is even, then $a=(n^2-4)/4$. A choice of generators is $(n/2)e_1+e_2$ for $G_1$ and $e_1$ for $G_2$.
\end{itemize}

Finally, after manipulating determinants like we did when computing $D'$ and $D$, we find
\begin{multline*}
    \left(K(N)^{-1}\right)_{ij}=\frac{1}{D(N)}[(-1)^{N-d}D'(d-1)\\
    +(-1)^d D'(N-d-1)],
\end{multline*}
where $d=|i-j|$. Eq.~\ref{eq:K_nF_inv} then follows by plugging in $n=3$.

\section{Determining $K$ matrix from statistics}
\label{sec:Km_statistics}

In this appendix, we answer the following question: given an abelian topological order with its anyon fusion and statistics specified, how does one construct a corresponding CS theory?

More precisely, the setup of the problem consists of:
\begin{enumerate}[leftmargin=*]
    \item A finite abelian fusion group $G$. We write the fusion product of $x$ and $y$ as $x+y$ instead of the usual $xy$.
    
    \item A symmetric bilinear function $b:G\times G\to\mathbb{Q}/\mathbb{Z}$ which gives the braiding statistic $e^{2\pi ib(x,y)}$ between anyons $x$ and $y$. Bilinearity means that $b(x+y,z)=b(x,z)+b(y,z)$ and similarly for the second argument.
    
    \item A function $q:G\to\mathbb{Q}/2\mathbb{Z}$ which is related to $b$ via \[b(x,y)=\frac{1}{2}\left(q(x+y)-q(x)-q(y)\right),\] and determines topological spins by $\theta_x=e^{i\pi q(x)}$. With respect to a minimal generating set $\{e_1,...,e_n\}$ of $G$, we can write $q$ as a matrix $(q)_{ij}$ where $q_{ii}=q(e_i)\in\mathbb{Q}/2\mathbb{Z}$ and $q_{ij}=b(e_i,e_j)\in\mathbb{Q}/\mathbb{Z}$ if $i\neq j$.
\end{enumerate}
Note that $b$ does not determine $q$ even though the converse is true. Indeed, $q(x)=b(x,x)\text{ mod }1$, but $q(x)$ itself is defined mod 2. This is the minus sign ambiguity in determining exchange statistic from braiding statistic. We focus on bosonic topological orders and assume modularity of the topological $S$-matrix, which in our language means that $(G,q)$ is \textit{non-degenerate} in the sense that if $b(x,y)=0$ for all $y$, then $x=0$. We comment on fermionic topological orders in Section~\ref{subsec:fermionic}.

Our goal is to find a $K$ matrix that produces the $(G,q)$ specified above. Naively, this is achieved by inverting the matrix $q$. For example, the toric code has \[q=
\begin{pmatrix}
0 & \frac{1}{2}\\[3pt]
\frac{1}{2} & 0
\end{pmatrix},\]
so we can take \[
K=q^{-1}=
\begin{pmatrix}
0 & 2\\
2 & 0
\end{pmatrix}.\]
However, $q^{-1}$ is not an integer matrix for a generic $q$. For example, the three-fermion theory has $G=\mathbb{Z}_2\times\mathbb{Z}_2$ and \[q=
\begin{pmatrix}
1 & \frac{1}{2}\\[3pt]
\frac{1}{2} & 1
\end{pmatrix},\]
but $q^{-1}$ is not an integer matrix. Instead, we need to ``enlarge'' $q$ to $\tilde{q}$ by adding transparent bosons (i.e.~bosons that braid trivially with everything) to the bottom right corner:
\[\tilde{q}=
\begin{pmatrix}
1 & \frac{1}{2} & 0 & 0\\[3pt]
\frac{1}{2} & 1 & 1 & 0\\[1.5pt]
0 & 1 & 2 & 1\\
0 & 0 & 1 & 2
\end{pmatrix}
\Rightarrow K=\tilde{q}^{-1}=
\begin{pmatrix}
4 & -6 & 4 & -2\\
-6 & 12 & -8 & 4\\
4 & -8 & 6 & -3\\
-2 & 4 & -3 & 2
\end{pmatrix}.\]
To obtain an enlargement algorithm that works for arbitrary $(G,q)$, we follow the strategy by Wall \cite{WALL1963281, wang2020abelian}: first we present a structure theorem for $(G,q)$, which classifies all irreducible building blocks of $q$ and gives an algorithm for decomposing $q$ into these blocks. Then we write down an enlargement for each irreducible block.

\subsection{Structure theorem}
Given $(G,q)$ and subgroups $G_1$, $G_2$ of $G$, we say that $G$ is the orthogonal direct product of $G_1$ and $G_2$ if $G=G_1\times G_2$ and $b(x_1,x_2)=0$ for all $x_1\in G_1$, $x_2\in G_2$. We have the following structure theorem:

\begin{nthm}
\label{thm}
If $(G,q)$ is non-degenerate, then $G$ can be written as an orthogonal direct product $G=\prod_i G_i$ such that $\left(G_i,q|_{G_i}\right)$ is in one of the following irreducible classes labelled by letters $A$ through $F$:
\begin{enumerate}[leftmargin=*]
    \item $A_{2^k}\cong\mathbb{Z}_{2^k}$, and $q=\left(2^{-k}\right)$.
    
    \item $A_{p^k}\cong\mathbb{Z}_{p^k}$, $p>2$ prime, and $q=\left(2\alpha p^{-k}\right)$ where $\alpha$ is coprime with $p$ and is a quadratic residue mod $p$. ($x$ is a quadratic residue mod $p$ if $x=y^2\text{ mod }p$ for some $y$.) Different choices of $\alpha$ lead to the same $q$ up to change of generator.
    
    \item $B_{2^k}\cong\mathbb{Z}_{2^k}$, and $q=\left(-2^{-k}\right)$.
    
    \item $B_{p^k}\cong\mathbb{Z}_{p^k}$, $p>2$ prime, and $q=\left(2\beta p^{-k}\right)$ where $\beta$ is coprime with $p$ and is not a quadratic residue mod $p$. Different choices of $\beta$ lead to the same $q$ up to change of generator.
    
    \item $C_{2^k}\cong\mathbb{Z}_{2^k}$, $k\ge2$, and $q=\left(5\times2^{-k}\right)$.
    
    \item $D_{2^k}\cong\mathbb{Z}_{2^k}$, $k\ge2$, and $q=\left(-5\times2^{-k}\right)$.
    
    \item $E_{2^k}\cong\mathbb{Z}_{2^k}\times\mathbb{Z}_{2^k}$, and $q=
    \begin{pmatrix}
    0 & 2^{-k}\\
    2^{-k} & 0
    \end{pmatrix}$.
    
    \item $F_{2^k}\cong\mathbb{Z}_{2^k}\times\mathbb{Z}_{2^k}$, and $q=
    \begin{pmatrix}
    2^{1-k} & 2^{-k}\\
    2^{-k} & 2^{1-k}
    \end{pmatrix}$.
\end{enumerate}
\end{nthm}
The above decomposition is not unique, e.g.~$A_{p^k}\times A_{p^k}=B_{p^k}\times B_{p^k}$ and $A_2\times A_2\times A_2=A_2\times E_2$. The toric code is in class $E_2$ and the three-fermion theory is in $F_2$.

Before we describe how the decomposition in Theorem~\ref{thm} can be performed, we state the following useful lemma:
\begin{nlemma}
\label{lemma}
Let $(G,q)$ be non-degenerate, $H$ a subgroup of $G$ such that $(H,q|_H)$ is non-degenerate. Then $G$ is the orthogonal direct product of $H$ and its orthogonal complement $H^\circ:=\{g\in G:b(g,h)=0\ \forall\ h\in H\}$, and $(H^\circ,q|_{H^\circ})$ is also non-degenerate.
\end{nlemma}

We perform the decomposition in Theorem~\ref{thm} using the following three steps:

\begin{itemize}[leftmargin=*]
    \item \textbf{Step~1}. We can uniquely decompose
    \begin{equation}
    \label{eq:sylow}
        G=\prod\limits_{p\text{ prime}}G_p,
    \end{equation}
    where $G_p$ is the unique Sylow $p$-subgroup of $G$. This product is always orthogonal.
    
    \item \textbf{Step~2}. Now we replace $G=G_p$ for fixed $p$. Let $p^r$ be the exponent of $G$, i.e.~the least common multiple of the orders of all elements in $G$. Write $G$ as a (non-orthogonal) direct product of a homogeneous subgroup $H$ of exponent $p^r$,i.e.~$H\cong\mathbb{Z}_{p^r}^m$ for some $m$, and another subgroup of smaller exponent. One can show that $(H,q|_H)$ is non-degenerate. Lemma~\ref{lemma} then gives $G=H\times H^\circ$ which is an orthogonal direct product, and $H^\circ$ has exponent smaller than $p^r$. Proceeding in this way, we can decompose $G$ into an orthogonal direct product of homogeneous subgroups.
    
    \item \textbf{Step~3}. Replace $G$ again by a homogeneous group of exponent $p^r$, $r\ge 1$. We look for $x\in G$ such that $p^r b(x,x)\in \mathbb{Z}_{p^r}$ is coprime with $p$. Such $x$ need not exist when $p=2$, but when it exists it is often easy to spot by inspection. However, for generality we present a more organized method (readers may skip this part and jump to Cases~3.1 \& 3.2 below). Consider the subgroup $G_0=\left\{g\in G:\text{ord}(g)\le p^{r-1}\right\}$ of $G$, where $\text{ord}(g)$ is the order of $g$, and write $[x]$ for the coset containing $x$ in $G/G_0$. Define a new bilinear function $b':(G/G_0)\times(G/G_0)\to\mathbb{Q}/\mathbb{Z}$ by \[b'\left([x],[y]\right)=p^{r-1}b(x,y)\in\mathbb{Q}/\mathbb{Z}.\] Now we look for some $[x]$ such that $pb'([x],[x])\in\mathbb{Z}_p$ is coprime with $p$. If $p\neq 2$, such $[x]$ always exists, and although we may still need an exhaustive search, this search is easier since $G/G_0$ has a smaller size than $G$. If $p=2$, such $[x]$ exists if and only if the $i$th diagonal element of $p^{r-1}q$ is nonzero for some $i$, in which case the generating element $[e_i]$ satisfies our requirement. Our next step depends on whether or not such $[x]$ was found:

\textbf{Case~3.1}. We found some $[x]\in G/G_0$ with $pb'([x],[x])$ coprime with $p$. Then $\left(\left<x\right>,q|_{\left<x\right>}\right)$ is non-degenerate, where $x\in[x]$ is an arbitrary coset representative and $\left<x\right>$ is the subgroup of $G$ generated by $x$. Lemma~\ref{lemma} then gives $G=\left<x\right>\times\left<x\right>^\circ$, and we go back to Step~3.

\textbf{Case~3.2} (occurs only if $p=2$). $b'([x],[x])=0$ for all $[x]\in G/G_0$. Pick some $x\in G$ of order $2^r$, e.g.~a generating element $x=e_i$. One can show that there exists $y\in G$ (not necessarily unique) such that $b'([x],[y])=1/2$. Let $x\in[x]$ and $y\in[y]$ be arbitrary coset representatives. Then $\left(\left<x,y\right>,q|_{\left<x,y\right>}\right)$ is non-degenerate, and $\left<x,y\right>\cong E_{2^r}$ or $F_{2^r}$. Again we apply Lemma~\ref{lemma} and then go back to Step~3.
\end{itemize}

Recursive application of the above steps leads to full decomposition of $(G,q)$.

\subsection{Enlargement algorithm}

Now we describe how to enlarge the matrix $q$ to $\tilde{q}$ such that $K=\tilde{q}^{-1}$ is an integer matrix with even diagonal, so that $K$ describes a bosonic CS theory. Using Theorem~\ref{thm}, we assume without loss of generality that $(G,q)$ is in one of the classes $A$ through $F$.

\begin{enumerate}
    \item $(G,q)\cong A_{2^r}$, $B_{2^r}$ or $E_{2^r}$. No enlargement is needed.
    
    \item $(G,q)\cong A_{p^r}$ or $B_{p^r}$ with $p>2$. Write $q=(np^{-r})$ for some $-p^r<n<p^r$. Then there exist $d_1$ even and $d_2$ odd such that $1=nd_1-p^rd_2$ and $0<d_2<|d_1|$. Next, choose $a_1$ even such that $a_1d_2$ is the closest even multiple of $d_2$ to $d_1$, and write $d_1=a_1d_2-d_3$. Continuing this algorithm, we obtain
    \begin{align*}
        1&=nd_1-p^rd_2\\
        d_1&=a_1d_2-d_3\\
        d_2&=a_2d_3-d_4\\
        &\cdots\\
        d_{k-1}&=a_{k-1}d_k-1\\
        d_k&=a_k
    \end{align*}
    where $a_jd_{j+1}$ is always the closest even multiple of $d_{j+1}$ to $d_j$. Then we take
    \setcounter{MaxMatrixCols}{6}
    \[
    \tilde{q}=
    \begin{pmatrix}
    np^{-r} & 1\\
    1&  a_1 & 1\\
    & 1 & a_2\\
    & & & \ddots\\
    & & & & a_{k-1} & 1\\
    & & & & 1 & a_k
    \end{pmatrix}.
    \]
    The algorithm we employed to produce $\{a_i\}$ is a variation of the \textit{Euclidean algorithm}.
    
    \item $(G,q)\cong C_{2^r}$ or $D_{2^r}$. In this case the Euclidean algorithm still works, but we need to take $d_1$ odd and $d_2$ even.
    
    \item $(G,q)\cong F_{2^r}$. We take
    \begin{equation}
        \label{eq:lift_F}
        \tilde{q}=
        \begin{pmatrix}
        2^{1-r} & 2^{-r} & &\\
        2^{-r} & 2^{1-r} & 1 &\\[1.5pt]
        & 1 & \frac{2}{3}(2^r+(-1)^{r-1}) & 1\\[1.5pt]
        & & 1 & 2(-1)^{r-1}
        \end{pmatrix}.
    \end{equation}
\end{enumerate}

\subsection{Example}
\label{subsec:lift_example}

We demonstrate the procedures in the previous sections with a coined example $G=\mathbb{Z}_8^5=\left<e_1,...,e_5\right>$, \[q=
\begin{pmatrix}
\frac{5}{8} & \frac{1}{4} & \frac{1}{8} & 0 & \frac{3}{8}\\[3pt]
\frac{1}{4} & \frac{5}{4} & 0 & \frac{7}{8} & \frac{1}{4}\\[3pt]
\frac{1}{8} & 0 & \frac{5}{8} & \frac{7}{8} & \frac{3}{4}\\[3pt]
0 & \frac{7}{8} & \frac{7}{8} & \frac{3}{2} & \frac{1}{2}\\[3pt]
\frac{3}{8} & \frac{1}{4} & \frac{3}{4} & \frac{1}{2} & \frac{7}{8}
\end{pmatrix}.\]
Since $G$ is already homogeneous, we jump straight to Step~3. First we spot that $e_1^Tqe_1=5/8$ has additive order 8 mod 1, so $\left(\left<e_1\right>,q|_{\left<e_1\right>}\right)$ is non-degenerate. We therefore compute $\left<e_1\right>^\circ=\left<f_1,f_2,f_3,f_4\right>$, where $f_1=-2e_1+e_2$, $f_2=e_1+3e_3$, $f_3=e_4$, $f_4=e_1+e_5$. With respect to these generators, we have \[q_1:=q|_{\left<e_1\right>^\circ}=
\begin{pmatrix}
\frac{3}{4} & \frac{1}{4} & \frac{7}{8} & \frac{1}{2}\\[3pt]
\frac{1}{4} & 1 & \frac{5}{8} & \frac{5}{8}\\[3pt]
\frac{7}{8} & \frac{5}{8} & \frac{3}{2} & \frac{1}{2}\\[3pt]
\frac{1}{2} & \frac{5}{8} & \frac{1}{2} & \frac{1}{4}
\end{pmatrix}.\]
Since all diagonal entries of $q_1$ have denominators at most 4, we turn to Case~3.2 and pick any generator, say $f_1$. The equation $f_1^Tq_1y=1/8$ has a solution $y=-f_3$. Then we work out \[\left<f_1,-f_3\right>^\circ=\left<-3f_1+f_2,4f_1+4f_3+f_4\right>.\] With respect to the generators $\{f_1,-f_3,-3f_1+f_2,4f_1+4f_3+f_4\}$, we have \[q_1=q_2\oplus q_3=
\begin{pmatrix}
\frac{3}{4} & \frac{1}{8}\\[3pt]
\frac{1}{8} & \frac{3}{2}
\end{pmatrix}
\oplus
\begin{pmatrix}
\frac{1}{4} & \frac{1}{8}\\[3pt]
\frac{1}{8} & \frac{1}{4}
\end{pmatrix},\]
where $\oplus$ is the direct sum of matrices on the direct product group. Picking appropriate generators $\{f_1+f_3,2f_1-3f_3\}$, we can put $q_2$ into a standard form \[q_2=
\begin{pmatrix}
0 & \frac{1}{8}\\[3pt]
\frac{1}{8} & 0
\end{pmatrix}.\]
To summarize,
\begin{equation}
    \label{eq:example_q}
    q\cong\left(\frac{5}{8}\right)\oplus
\begin{pmatrix}
0 & \frac{1}{8}\\[3pt]
\frac{1}{8} & 0
\end{pmatrix}
\oplus
\begin{pmatrix}
\frac{1}{4} & \frac{1}{8}\\[3pt]
\frac{1}{8} & \frac{1}{4}
\end{pmatrix}
\cong C_8\times E_8\times F_8
\end{equation}
with respect to the generators \[\{e_1,f_1+f_3,2f_1-3f_3,-3f_1+f_2,4f_1+4f_3+f_4\}.\] However, this decomposition is not unique; with respect to some other generators, we also have \[q\cong \left(\frac{5}{8}\right)\oplus\left(-\frac{1}{8}\right)\oplus\left(-\frac{1}{8}\right)\oplus\left(-\frac{1}{8}\right)\oplus\left(-\frac{5}{8}\right).\]
Next, we enlarge each summand in Eq.~\ref{eq:example_q}. To enlarge the $C_8$, we apply the Euclidean algorithm:
\begin{align*}
1&=5\times13-8\times8\\
13&=2\times8-3\\
8&=2\times3-(-2)\\
3&=(-2)\times(-2)-1\\
-2&=(-2)\times1
\end{align*}
which gives \[\left(\frac{5}{8}\right)\mapsto
\begin{pmatrix}
\frac{5}{8} & 1\\[1.5pt]
1 & 2 & 1\\
& 1 & 2 & 1\\
& & 1 & -2 & 1\\
& & & 1 & -2
\end{pmatrix}.\]
The $E_8$ does not need enlargement, and the $F_8$ can be enlarged to Eq.~\ref{eq:lift_F}.

This completes our example. The total inverse $K$ matrix is
\begin{equation*}
    K^{-1}=\begin{pmatrix}
    \frac{5}{8} & 1\\[1.5pt]
    1 & 2 & 1\\
    & 1 & 2 & 1\\
    & & 1 & -2 & 1\\
    & & & 1 & -2
    \end{pmatrix}
    \oplus
    \begin{pmatrix}
    0 & \frac{1}{8}\\[3pt]
    \frac{1}{8} & 0
    \end{pmatrix}
    \oplus
    \begin{pmatrix}
    \frac{1}{4} & \frac{1}{8}\\[3pt]
    \frac{1}{8} & \frac{1}{4} & 1\\[1.5pt]
    & 1 & 6 & 1\\
    & & 1 & 2
    \end{pmatrix},
\end{equation*}
and the total $K$ matrix is
\begin{align*}
    K=&\begin{pmatrix}
    104 & -64 & 24 & 16 & 8\\
    -64 & 40 & -15 & -10 & 5\\
    24 & -15 & 6 & 4 & 2\\
    16 & -10 & 4 & 2 & 1\\
    8 & -5 & 2 & 1 & 0
    \end{pmatrix}
    \oplus
    \begin{pmatrix}
    0 & 8\\
    8 & 0
    \end{pmatrix}\\
    &\qquad\oplus
    \begin{pmatrix}
    48 & -88 & 16 & -8\\
    -88 & 176 & -32 & 16\\
    16 & -32 & 6 & -3\\
    -8 & 16 & -3 & 2
    \end{pmatrix}.
\end{align*}

\subsection{Fermionic case}
\label{subsec:fermionic}

Finally, we consider fermionic topological orders. Now a local fermion excitation is a superselection sector that braids trivially with everything, i.e.~a transparent fermion, so we need to modify our non-degeneracy assupmtion. We assume that $(G,q)$ is \textit{weakly non-degenerate} in the sense that if $b(x,y)=0$ for all $y$ and $q(x)=0$, then $x=0$.

Suppose that $\psi$ is a transparent fermion. Since $2\psi=0$, in the decomposition Eq.~\ref{eq:sylow} we must have $\psi\in G_2$. Suppose $\psi=mx$ for some $m\in\mathbb{Z}$ and $x\in G_2$. Then
\begin{align*}
    0=b(x,2\psi)=2mb(x,x)&\text{ mod }2,\\
    1=q(\psi)=m^2b(x,x)&\text{ mod }2,
\end{align*}
so $m$ must be odd. But $2\psi=2mx=0\in G_2$, so $2x=0$ and hence $\psi=x$. Thus we have proved that $\psi$ is not a non-trivial multiple of any $x$, so $G_2$ can be decomposed into an orthogonal direct product of $\left<\psi\right>=\{0,\psi\}$ and $\left<\psi\right>^\circ$. Continuing this process, we end up with $G=\mathbb{Z}_2^r\times G'$ where each $\mathbb{Z}_2$ is generated by a transparent fermion and $(G',q|_{G'})$ is non-degenerate. The bosonic result can then be applied to $(G',q|_{G'})$.

As an example, consider the $\nu=1/3$ fractional quantum Hall effect. Treated as a bosonic theory, the fusions group is $G=\mathbb{Z}_6$, whose generator we call $x$, and $q=(1/3)$. This theory is only weakly non-degenerate, with $3x$ a transparent fermion. Following the above recipe, we decompose $G=\mathbb{Z}_2\times\mathbb{Z}_3=\left<3x\right>\times\left<2x\right>$. Now $\left(\left<2x\right>,q|_{\left<2x\right>}\right)$ is non-degenerate, where
\begin{equation*}
    q|_{\left<2x\right>}=\left(\frac{4}{3}\right)=\left(-\frac{2}{3}\right).
\end{equation*}
We can then use the Euclidean algorithm to enlarge it to
\begin{equation*}
    \tilde{q}=
    \begin{pmatrix}
    -\frac{2}{3} & 1\\[1.5pt]
    1 & -2
    \end{pmatrix}
    \implies K=\tilde{q}^{-1}=
    \begin{pmatrix}
    -6 & -3\\
    -3 & -2
    \end{pmatrix}.
\end{equation*}
Putting the transparent fermion back, we get a $3\times 3$ matrix which can be mapped through a general linear transformation as follows:
\begin{equation*}
    W
    \begin{pmatrix}
    -6 & -3\\
    -3 & -2\\
    & & 1
    \end{pmatrix}
    W^T=
    \begin{pmatrix}
    3\\
    & -1\\
    & & -1
    \end{pmatrix},
\end{equation*}
where
\begin{equation*}
    W=
    \begin{pmatrix}
    1 & 0 & 3\\
    0 & 1 & 1\\
    -1 & 1 & -1
    \end{pmatrix}.
\end{equation*}
This shows the equivalence of our result with the standard $K$ matrix (3) for the $\nu=1/3$ fractional quantum Hall state.

\end{document}